\documentclass[%
 reprint,
superscriptaddress,
 amsmath,amssymb,
 aps,
]{revtex4-2}

\usepackage{graphicx}
\usepackage{dcolumn}
\usepackage{bm}
\usepackage{braket}

\begin{document}


\title{Improving the Performance of Digitized Counterdiabatic Quantum Optimization via Algorithm-Oriented Qubit Mapping}

\author{Yanjun Ji}
\email{quantumji@hotmail.com}
\affiliation{Institute of Computer Architecture and Computer Engineering, University of Stuttgart, Pfaffenwaldring 47, 70569 Stuttgart, Germany}

\author{Kathrin F. Koenig}
\affiliation{Fraunhofer Institute for Applied Solid State Physics IAF, Tullastrasse 72, 79108 Freiburg, Germany}
\affiliation{Faculty of Engineering, University of Freiburg, Georges-Köhler-Allee 101, 79110 Freiburg, Germany}

\author{Ilia Polian}
\email{ilia.polian@informatik.uni-stuttgart.de}
\affiliation{Institute of Computer Architecture and Computer Engineering, University of Stuttgart, Pfaffenwaldring 47, 70569 Stuttgart, Germany}

\date{\today}

\begin{abstract}

This paper presents strategies to improve the performance of digitized counterdiabatic quantum optimization algorithms by cooptimizing gate sequences, algorithm parameters, and qubit mapping. Demonstrations on real hardware validate the effectiveness of these strategies, leveraging both algorithmic and hardware advantages.Specifically, our approach achieves an average increase in approximation ratio of 4.49$\times$ without error mitigation and 84.8\% with error mitigation compared to Qiskit and Tket on IBM quantum processors. These findings provide valuable insights into the codesign of algorithm implementation, tailored to optimize qubit mapping and algorithm parameters, with broader implications for enhancing algorithm performance on near-term quantum devices.

\end{abstract}

\maketitle


\section{\label{sec:intro}Introduction}

Quantum computing offers the potential to solve complex problems that are difficult for classical computers. However, current quantum processors, operating in the noisy intermediate-scale quantum (NISQ) era, face inherent limitations in scale and operation \cite{preskill2018quantum}. NISQ devices typically comprise several tens to hundreds of qubits, and their qubit connectivity is restricted, with each qubit capable of directly interacting with only a few neighboring qubits \cite{jurcevic2021demonstration}. Moreover, noise and errors impose constraints on the reliability and complexity of algorithms executed on such devices, posing significant challenges for implementing quantum algorithms effectively.
To tackle these challenges, researchers have proposed various compilation and optimization techniques \cite{shi2019optimized, gokhale2020optimized, jones2022robust, ji2022calibration, earnest2021pulse, ji2023optimizing} to transform quantum circuits to satisfy qubit connectivity constraints while optimizing them to reduce errors and enhance performance. Meanwhile, hybrid quantum-classical algorithms such as the quantum approximate optimization algorithm (QAOA) \cite{farhi2014quantum} have emerged as a promising approach for near-term quantum applications \cite{moll2018quantum}.

The QAOA is a prominent variational algorithm \cite{cerezo2021variational} that leverages iterative classical and quantum steps to approximate solutions for combinatorial optimization problems. To enhance its efficiency, numerous variants \cite{blekos2024review} have been proposed, including adaptive parameter updates \cite{zhu2022adaptive}, digitized versions \cite{chandarana2022digitized, chandarana2022digitized, chai2022shortcuts, wurtz2022counterdiabaticity}, and the recent work on mean-field optimization \cite{misra2023mean}. Here we focus on the digitized counterdiabatic QAOA (DC-QAOA) \cite{chandarana2022digitized, chai2022shortcuts, wurtz2022counterdiabaticity}, which incorporates an additional driving Hamiltonian inspired by quantum shortcuts to adiabaticity \cite{xi2010shortcut, odelin2019shortcuts}. This additional Hamiltonian leads to faster convergence to the ground state energy, allowing for a reduction in the number of depths. DC-QAOA has demonstrated improved performance in various applications, such as portfolio optimization \cite{hegade2022portfolio} and protein folding problems \cite{chandarana2022digitizedpf}. However, high-quality compilation is crucial for efficient implementation of DC-QAOA on NISQ devices.

Qubit mapping is a crucial task in executing quantum algorithms on near term quantum devices, which involves mapping logical qubits in a quantum circuit to physical qubits on the device, taking into account the device's connectivity constraints and noise \cite{li2019tackling, cowtan2019qubit, zhu2022an, niu2020hardware, zhang2021time}. 
The quality of qubit mapping significantly impacts the performance of algorithms, as it may necessitate the insertion of additional SWAP gates to implement the logical connectivity of the circuit, and a SWAP gate is typically implemented with three CX gates, which are a major source of errors in quantum devices. Furthermore, it is essential to consider the time varying errors on quantum devices for performance improvement \cite{ji2022calibration}.

Various methods have been proposed to address the qubit mapping problem, offering different trade-offs between solution quality and computational complexity. While exact methods \cite{bhattacharjee2017depth, ShafaeiSP14, murali2019noise, siraichi2019qubit, tan2020optimal} guarantee optimal solutions by minimizing the number of SWAP gates, circuit depth, or maximizing circuit fidelity, they are computationally demanding and may not scale well for large circuits or devices. Heuristic methods \cite{siraichi2019qubit, zulehner2018efficient, childs19, li2020qubits, niu2020hardware, murali2019noise}, on the other hand, aim to find near optimal solutions within a reasonable time frame, but they do not guarantee the best solution, and their effectiveness can vary depending on the specific circuit or device. Another approach to address the qubit mapping problem is the use of swap layers or strategies \cite{ji2023algorithmoriented, harrigan2021quantum, kivlichan2018quantum, o2019generalized, weidenfeller2022scaling, hashim22optimized}, which are predefined sequences of SWAP gates applicable to specific circuits. Swap strategies provide scalable solutions for handling large quantum circuits or devices, but may not guarantee optimality.

In recent developments, algorithm-oriented qubit mapping (AOQMAP) \cite{ji2023algorithmoriented} has emerged as an efficient approach for mapping variational quantum algorithms (VQAs), with a primary focus on achieving both optimality and scalability. This approach follows a two-step methodology. In the first step, AOQMAP generates executable quantum circuits tailored to the specific target quantum device, considering the connectivity constraints of subtopologies and decomposing them into basis gates of the target quantum processing unit (QPU). Unlike conventional approaches that map the entire circuit, AOQMAP adopts a strategy of solely mapping two-qubit interactions, leveraging the flexibility of applying single qubit gates to any qubit. This approach provides optimal and scalable solutions for VQAs featuring all-to-all connected two-qubit interactions on linear, T-shaped, and H-shaped subtopologies commonly found in various NISQ devices, such as Google's Sycamore \cite{arute2019quantum}, IBM's QPUs \cite{corcoles2019challenges}, and Rigetti's processors \cite{otterbach2017unsupervised}. In the case of partially connected two-qubit interactions, AOQMAP provides solutions by optimizing the initial qubit order to minimize additional CX gates. In the second step, subtopology-adapted circuits are mapped onto the target quantum device by matching subtopologies and selecting the configuration that minimizes the circuit error according to a predefined cost function. This mapping step incorporates calibration data to account for noise characteristics of the target QPU, ensuring minimal impact from noise during the execution.

AOQMAP distinguishes itself by initiating mapping directly from the algorithm's two-qubit Hamiltonian, which allows for optimization opportunities during the process of transforming into a circuit implementation. This codesign approach simultaneously satisfies algorithmic requirements and hardware constraints. It also facilitates pulse level optimization, improving efficiency \cite{ji2023optimizing}. Additionally, the results obtained through AOQMAP can be easily transferred across various QPUs and algorithms that share similar structures, providing flexibility and practical utility in developing quantum algorithms.

In this paper, we delve into techniques to enhance the performance of DC-QAOA by cooptimizing qubit mapping alongside algorithm parameters and structure.
We demonstrate that AOQMAP \cite{ji2023algorithmoriented} achieves significant reductions in CX gate count and circuit depth compared to both general-purpose compilers (e.g., Qiskit \cite{qiskit}, Tket \cite{sivarajah2020tket}) and application-specific compilers (e.g., 2QAN \cite{lao20222qan}). In comparison to general-purpose compilers, it reduces CX gates by up to 77.2\% and circuit depth by up to 90.8\%. Furthermore, AOQMAP surpasses the application-specific compiler \cite{lao20222qan} by reducing CX gates by up to 16.3\% and circuit depth by up to 84.9\%. Demonstrations on various IBM QPUs show that the integration of AOQMAP with DC-QAOA yields significant advantages, including the generation of high-quality solutions for qubit mapping and the augmentation of performance at algorithmic level. Specifically, compared to Qiskit and Tket, AOQMAP achieves an average reduction of 28.8\% in CX gate count and 33.4\% in circuit depth for benchmarked problem instances. Moreover, AOQMAP enhances approximation ratio by $4.49 \times$ in the absence of error mitigation and 84.8\% in the presence of error mitigation techniques. These findings provide valuable insights into the realm of algorithm-hardware codesign and demonstrate efficacy of cooptimizing qubit mapping with the algorithm itself.

The remainder of this paper is structured as follows. Section \ref{sec:method} details the methodology for applying the AOQMAP to DC-QAOA with various formulations and presents optimization strategies to enhance algorithm performance.
Section~\ref{sec:applic} showcases the effectiveness of our optimization method through its application to diverse problems, including MaxCut on complete and noncomplete graphs, as well as portfolio optimization.
Section \ref{sec:bench} presents comprehensive benchmarking experiments and results, encompassing evaluations of qubit mapping strategies,  analyses of noise influence and error mitigation techniques, and demonstrations on IBM QPUs. Finally, Sec. \ref{sec:concl} discusses results and concludes. The specifications of the IBM QPUs and error mitigation strategies used in this study are provided in Appendixes~\ref{app:cloud_platf} and \ref{app:error_mitigation}, respectively.

\section{\label{sec:method}Methodology}

This section delves into strategies for enhancing the performance of DC-QAOA while adhering to hardware constraints. We demonstrate their effectiveness through three scenarios: MaxCut on complete graphs, MaxCut on non-complete graphs, and portfolio optimization.

To illustrate these strategies, we employ the MaxCut problem on complete graphs as our problem instance. The MaxCut problem is a fundamental optimization problem in computer science that involves partitioning the nodes of a graph into two disjoint sets in a way that maximizes the number of edges crossing between the two sets. Due to its computational complexity, MaxCut is classified as NP-hard. The Hamiltonian for the MaxCut problem is formulated as
\begin{equation}
H_c = \frac{1}{2} \sum_{i,j} (1 - Z_i Z_j),
\end{equation}
where $Z_i$ is the Pauli $Z$ operator acting on qubit $i$. We use the standard mixer Hamiltonian \cite{farhi2014quantum}
\begin{equation}
H_m = \sum_{i} X_i,
\label{eq:mixer}
\end{equation}
where $X_i$ represents the Pauli $X$ operator acting on the qubit $i$.
The counterdiabatic (CD) driving Hamiltonian $H_{cd}$ is derived from adiabatic gauge potentials \cite{kolodrubetz2017geometry, chandarana2022digitized, chai2022shortcuts}.
In this work, we investigate two forms of the CD driving term: $\sum_{i,j} Z_iY_j$ and $\frac{1}{2}\sum_{i,j} (Z_iY_j+Y_iZ_j)$. The evolution of DC-QAOA with depth $p$ is given by
\begin{equation}
\ket{\psi_f} = U_{\text{DC}}(\gamma_p,\beta_p,\alpha_p) \ldots U_{\text{DC}}(\gamma_1,\beta_1,\alpha_1) \ket{\psi_0},
\end{equation}
where $\ket{\psi_0}$ denotes the ground state of $H_m$ and $U_{\text{DC}}(\gamma,\beta,\alpha) = U_{cd}(\gamma)U_m(\beta)U_{c}(\alpha)$ with $U_c(\alpha) = e^{-i\alpha H_c}$, $U_m(\beta) = e^{-i\beta H_m}$, and $U_{cd}(\gamma) = e^{-i\gamma H_{cd}}$.

The objective of DC-QAOA is to identify a set of angles that minimizes the expectation value of the problem Hamiltonian $\bra{\psi_f} H_c \ket{\psi_f}$ using classical optimization algorithms, such as gradient-based optimizers \cite{ahmadianfar2020gradient} or gradient-free optimizers like the constrained optimization by linear approximation (COBYLA) \cite{powell1994direct}. These optimizers iteratively adjust the angles to approximate the ground state of $H_c$, thereby providing an approximate solution to the MaxCut problem.
The approximation ratio of QAOA or DC-QAOA for MaxCut is defined as
\begin{equation}
    r = E/E_{0},
\end{equation}
where $E$ denotes the expectation value of the problem Hamiltonian obtained through QAOA or DC-QAOA, and $E_0$ represents the ground state energy of the system.

\subsection{Applying AOQMAP to DC-QAOA\label{subsec:apply_aoqmap}}

The Suzuki-Trotter decomposition \cite{suzuki1976generalized, suzuki1985decomposition, suzuki1976relationship} is a widely used method for approximating unitary time evolution operators. In the context of DC-QAOA, it is employed to approximate the adiabatic evolution of a system. This decomposition implies that the gate sequence for a particular Trotter step remains constant. Additionally, optimizing gate sequences is critical for actually simulating the system evaluation \cite{mansuroglu2023problem}. To satisfy the decomposition requirements, AOQMAP employs a fixed gate sequence for a specified depth in all instances of DC-QAOA. We explore four variations of DC-QAOA: DC-QAOA with a ZY CD driving term both with and without the problem Hamiltonian; and DC-QAOA with a ZY-YZ CD driving term both with and without the problem Hamiltonian.

\begin{figure*}
\centering
	\includegraphics[width=\linewidth]{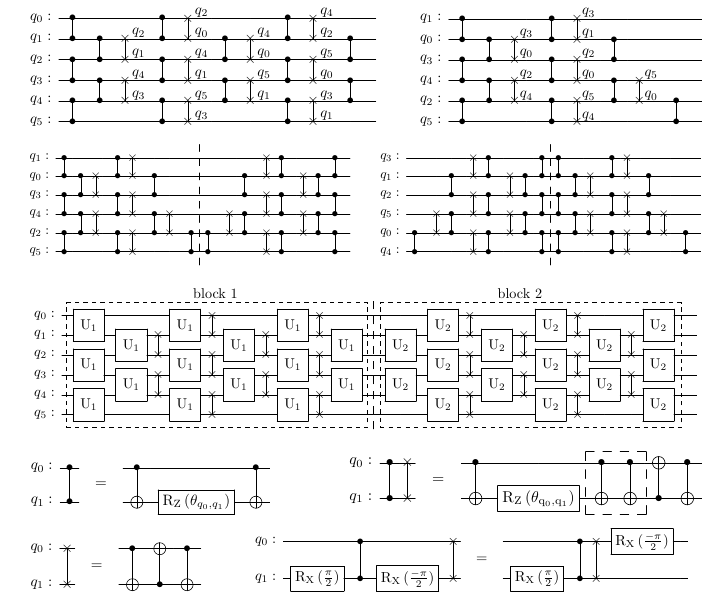}
        \put(-508,420){\textbf{(a)}}
	\put(-228,420){\textbf{(b)}}
	\put(-508,320){\textbf{(c)}}
	\put(-255,320){\textbf{(d)}}
	\put(-508,220){\textbf{(e)}}
	\put(-508,105){\textbf{(f)}}
	\put(-508,45){\textbf{(g)}}
        \put(-280,105){\textbf{(h)}}
	\put(-345,45){\textbf{(i)}}
	\caption{AOQMAP approach for two-qubit Hamiltonian in QAOA and DC-QAOA on a linear topology. (a) Routing solution for all-to-all connected ZZ gates in six-qubit QAOA at $p=1$. While an arbitrary initial qubit order allows for the execution of all two-qubit gates, it results in different gate sequences. (b) Routing solution for partially connected ZZ gates in six-qubit QAOA at $p =1$. The initial qubit order is optimized to minimize the total increase in the number of CX gates due to the insertion of SWAP gates. The mirror symmetry of swap layers is used to construct a solution for the next QAOA depth. Two constructions of mirror symmetry with varying initial mappings are displayed in (c) and (d); their corresponding gate sequences are referred to as AOQ-FS and AOQ-SF, respectively. (e) Solution for DC-QAOA leveraging mirror symmetry. The two-qubit gates in DC-QAOA are represented by $\mathrm{U_1}$ and $\mathrm{U_2}$ in blocks 1 and 2, respectively. Decomposition of (f) ZZ, (g) SWAP, and (h) ZZ-SWAP. Each SWAP gate adjacent to a ZZ gate introduces only one additional CX gate due to CX gate cancellation. (i) Optimization of ZY-SWAP gate. A ZY gate followed by a SWAP gate can be optimized by operating the SWAP gate directly behind the ZZ gate and altering the position of $\mathrm{R_x(\frac{-\pi}{2})}$ gate accordingly.}\label{fig:aoqmap_approach}
\end{figure*}
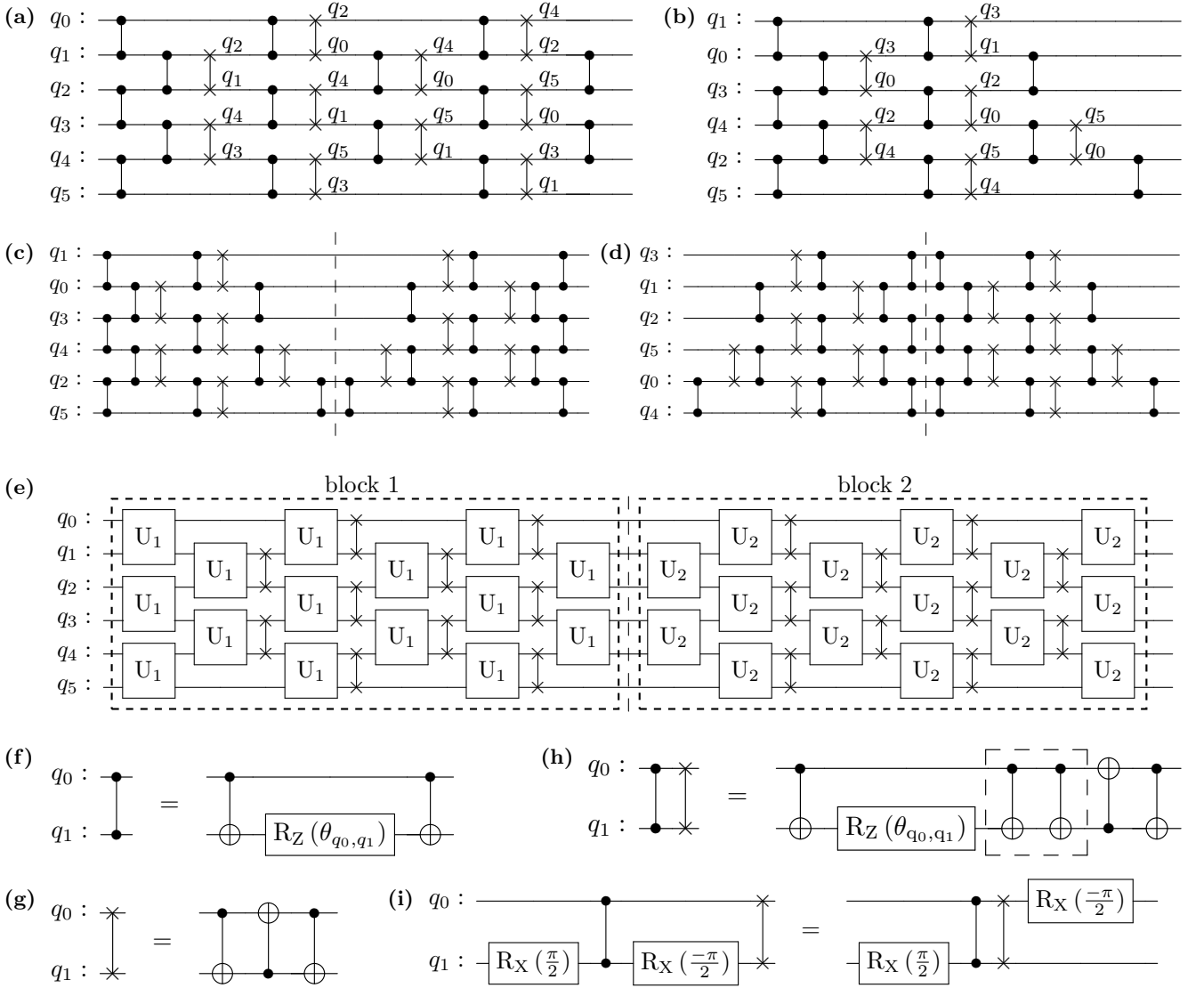

The routing solutions for QAOA and DC-QAOA with the AOQMAP approach on a linear topology are illustrated in Fig.~\ref{fig:aoqmap_approach}. Figure~\ref{fig:aoqmap_approach}(a) depicts the routing solution for fully connected ZZ gates in an $n$-qubit QAOA circuit. Each layer of ZZ gates is followed by a SWAP layer, except for the first and last layers, where the SWAP gate can be removed by adjusting the initial qubit order and the measurement order. Note that ZZ and SWAP gates are both undirected and commute, making the ZZ-SWAP gate equivalent to the SWAP-ZZ gate.
For QAOA with partially connected ZZ gates, the solution is obtained by optimizing the initial qubit order such that all existing ZZ gates are arranged as forwards as possible, and the end-located SWAP gates can be removed by adjusting the measurement order \cite{ji2023algorithmoriented}. An example of QAOA for MaxCut on noncomplete graphs is presented in Fig.~\ref{fig:aoqmap_approach}(b). In comparison to Fig.~\ref{fig:aoqmap_approach}(a), where an arbitrary initial qubit order enables the implementation of all ZZ gates, Fig.~\ref{fig:aoqmap_approach}(b) requires a specific initial qubit order to minimize the additional CX gate count. However, this deterministic initial qubit order is not unique, allowing for exploring different ZZ gate sequences that may result in distinct efficiencies of optimization performance, which we discuss later.

To construct the solution for DC-QAOA, we leverage the mirror symmetry of swap layers, which enables the generation of two distinct gate sequences, denoted as AOQ-FS and AOQ-SF, as illustrated in Figs.~\ref{fig:aoqmap_approach}(c) and \ref{fig:aoqmap_approach}(d), respectively. These two gate sequences are equivalent for all-to-all commuting gates such as ZZ gates but exhibit differences for non-all-to-all commuting gates such as ZY gates.
Figure~\ref{fig:aoqmap_approach}(e) presents general solutions for applying AOQMAP to DC-QAOA. The two-qubit gates in problem and/or CD driving Hamiltonians are represented by $\mathrm{U_1}$ and $\mathrm{U_2}$. For instance, in the case of DC-QAOA with a ZY CD driving term including problem Hamiltonian, $\mathrm{U_1}$ and $\mathrm{U_2}$ correspond to ZZ and ZY gates with respective parameters. Figures~\ref{fig:aoqmap_approach}(f)-(i) depict the decomposition of ZZ, SWAP, ZZ-SWAP, and ZY-SWAP gates, respectively. In particular, Fig.~\ref{fig:aoqmap_approach}(h) highlights that a SWAP gate directly following a ZZ gate introduces one additional CX gate due to CX gate cancellation, as depicted in the enclosed box. To optimize the SWAP gate following a ZY gate, this SWAP gate is repositioned before the final single qubit gate within the ZY gate, with the position of the corresponding single qubit gate adjusted accordingly, as illustrated in Fig.~\ref{fig:aoqmap_approach}(i). This optimization reduces CX gates involved.

The use of mirror symmetry in AOQMAP presents a distinctive characteristic, where the qubit order reverts to its initial state every two two-qubit Hamiltonian blocks. Specifically, when a depth-one circuit comprises an odd number of two-qubit Hamiltonian blocks, the final qubit order returns to its initial state until depth two. Although the Suzuki-Trotter decomposition mandates a fixed gate sequence for each Trotter step, variations in gate sequences within a single Trotter step can result in distinct Trotter errors. However, for a Hamiltonian composed solely of ZZ gates, gate sequences have no impact on Trotter error since ZZ gates commute with each other.

In the case of DC-QAOA with a ZY CD driving term but lacking problem Hamiltonian, maintaining a fixed gate sequence for each depth can be challenging. One potential solution is to employ the mirror symmetry of swap layers after each depth, reversing the final qubit mapping back to the initial one. However, this approach introduces a significant increase in noise. An alternative strategy is to utilize these swap layers to construct a solution that incorporates per layer reversed ZY gates at depth two. The utilization of reversed gate sequences has the potential to mitigate Trotter error \cite{childs2019faster, faehrmann2022randomizing, childs2021theory}, making AOQMAP a promising approach. In the case where problem Hamiltonian is included, ZZ and ZY terms at depth one reverse the final qubit mapping directly to the initial one, resulting in a fixed gate sequence for each depth.

For DC-QAOA with a ZY-YZ CD driving term but excluding problem Hamiltonian, ZY and YZ terms maintain a fixed gate sequence at each depth. However, when problem Hamiltonian is included, DC-QAOA involves three terms: ZZ, ZY, and YZ. Alternating between swap layers and their mirror symmetry ensures a consistent gate sequence every two depths.

In summary, by exploiting swap layers and their symmetry, the AOQMAP strategy ensures an efficient construction of DC-QAOA circuits adhering to the noncommutative nature of CD driving terms. This approach results in a fixed gate sequence that depends on the initial qubit order. While the choice of gate sequence has no impact on the performance when gates within each block commute, the effectiveness of the parameter optimization process may vary when not all gates commute, which we discuss in the following section.

\subsection{Optimization strategies for improved algorithm performance}
\label{subsec:opt_strategies}

This section explores optimization strategies to enhance the performance of DC-QAOA with AOQMAP. The integration of SWAP gates during the qubit mapping process generates a deterministic gate sequence. In contrast, conventional DC-QAOA implementation does not impose a specific gate sequence and follows a sequential order where qubit 0 is successively connected to remaining qubits, followed by the sequential connection of qubits 1, 2, and so on, until reaching the final qubit $n$. We refer to this gate sequence as the ORIG sequence.
Although the performance of DC-QAOA generally remains robust to variations in Hamiltonian and gate sequence within a certain range of Trotter error, these factors can influence the algorithm's optimization efficiency, presenting opportunities for further performance enhancements.

\subsubsection{Iterations and initial guess values}

The gradient descent optimization method is used to minimize the expectation value of the problem Hamiltonian, corresponding to the objective function. This study investigates the influence of two critical factors during the optimization process: the maximum number of iterations and the number of random initial guess values.

\begin{figure}
\centering
	\includegraphics[width=\linewidth]{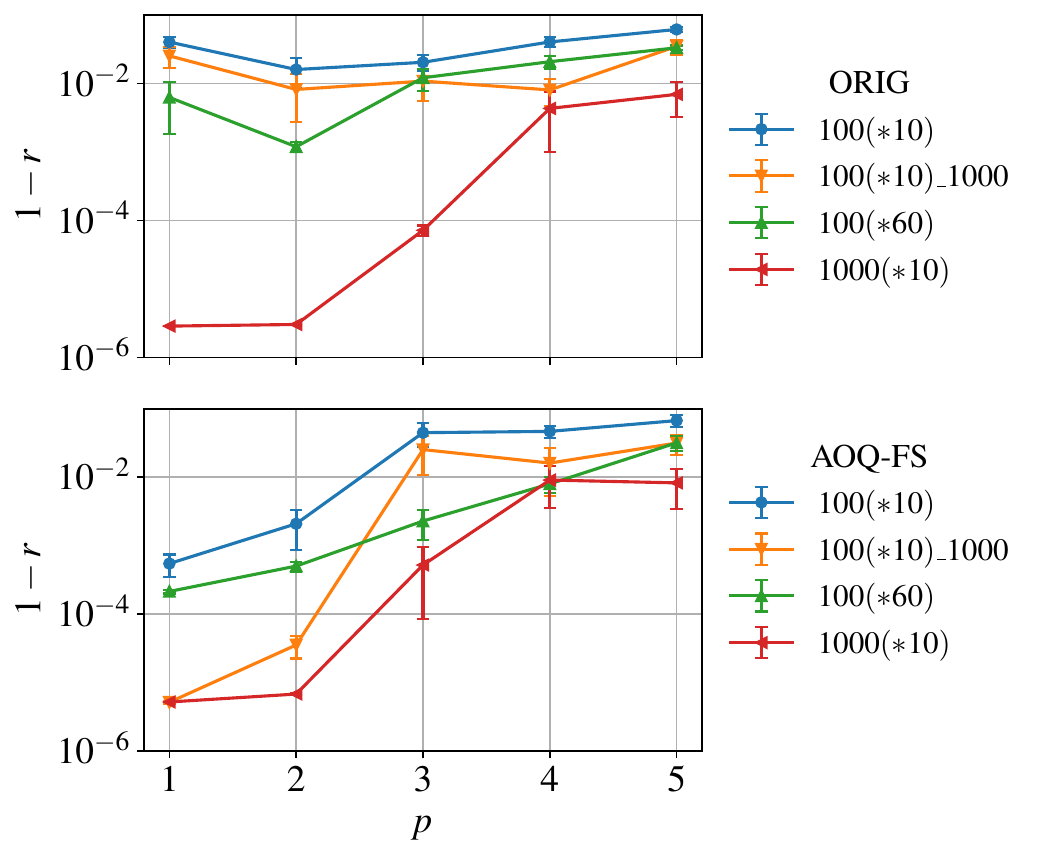}
 \put(-\linewidth,0.8\linewidth){\textbf{(a)}}
 \put(-\linewidth,0.43\linewidth){\textbf{(b)}}
	\caption{Comparison of different settings of the maximum number of iterations and the number of random initial guess values in the optimization process for the six-qubit DC-QAOA with a ZY-ZZ-X Hamiltonian sequence. Deviation $1-r$ of approximation ratio for two different gate sequences: (a) ORIG and (b) AOQ-FS. A lower value of $1-r$ indicates a higher approximation ratio, signifying improved performance. All error bars represent one standard error of the mean (SEM) across 10 independent repetitions.}\label{fig:compare_iterations}
\end{figure}

Figure~\ref{fig:compare_iterations} compares different settings for the maximum iteration and the number of random initial guess values in the optimization process. The first setting, denoted as 100(*10), uses 100 iterations and 10 random initial guess values. Each data point is repeated 10 times, and the error bar represents one standard error of the mean (SEM). The second setting, denoted as 100(*10)\_1000, maintains 100 iterations and 10 random guess values and increases the number of iterations for the optimized initial guess value to 1000, improving both ORIG and AOQ-FS gate sequences. In addition, AOQ-FS shows significantly high approximation ratios at lower depths. However, optimizing DC-QAOA becomes more difficult at larger depths, resulting in lower approximation ratios.
Increasing the number of random initial guess values to 60, represented as 100(*60), leads to improved performance compared to 100(*10). The best performance is observed when setting the maximum number of iterations to 1000 and the number of initial guess values to 10, yielding the highest approximation ratio for both ORIG and AOQ-FS gate sequences.

The results demonstrate that despite sharing the same order of Trotter errors, different gate sequences significantly influence optimization efficiency. The AOQ-FS gate sequence generally outperforms the ORIG gate sequence for these settings. To minimize the impact of the optimization on algorithm performance, we use the setting 1000(*10) during the optimization process.

\subsubsection{Hamiltonian and gate sequences}

\begin{figure*}[t]
\centering
	\includegraphics[width=\linewidth]{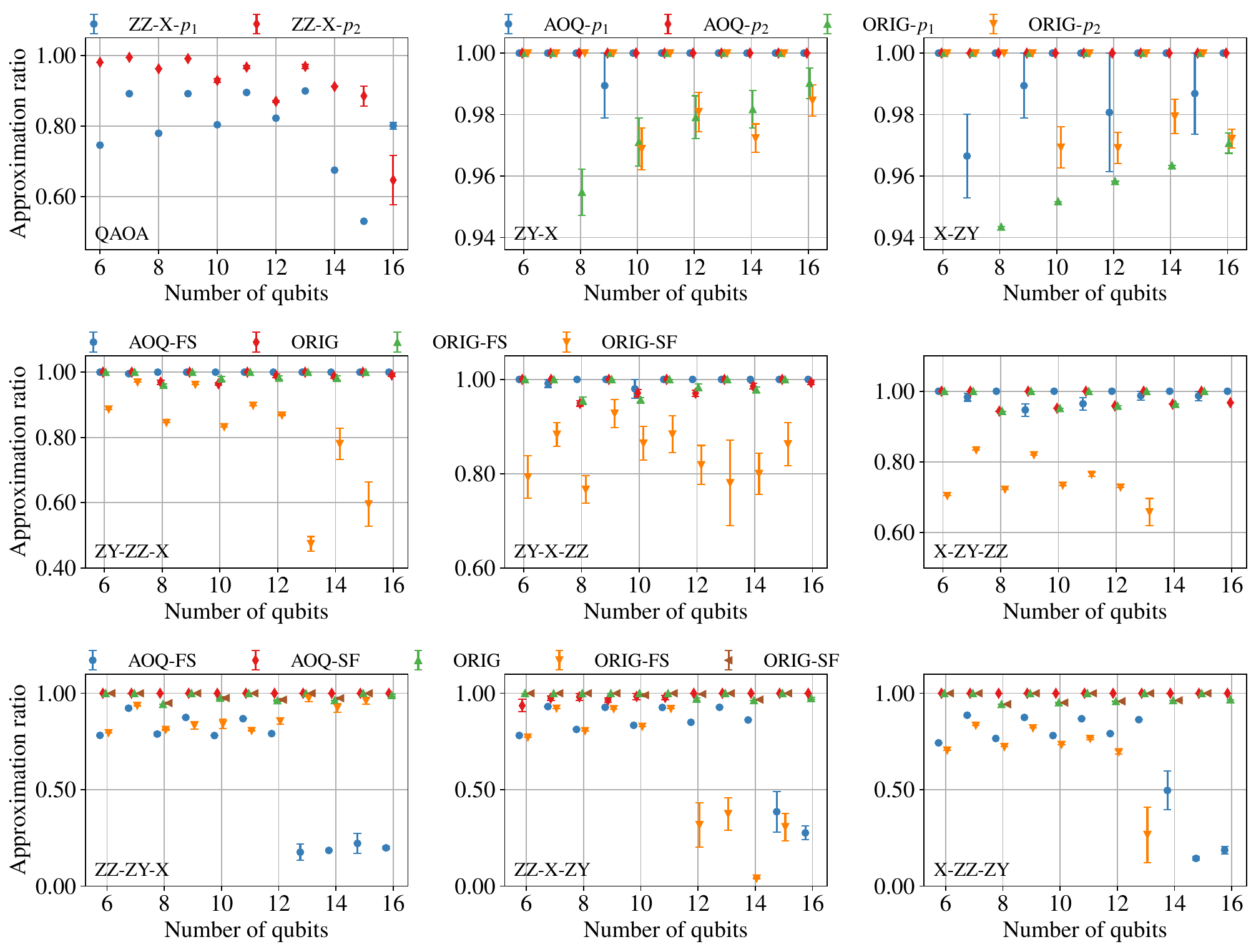}
 \put(-\linewidth,0.73\linewidth){\textbf{(a)}}
 \put(-\linewidth,0.48\linewidth){\textbf{(b)}}
 \put(-\linewidth,0.23\linewidth){\textbf{(c)}}
	\caption{Approximation ratio of QAOA and DC-QAOA with a ZY CD driving term for MaxCut on complete graphs. The Hamiltonian sequences differ depending on the order in which they operate on the initial state. For instance, ZY-X indicates that DC-QAOA, excluding problem Hamiltonian, applies the CD driving term first, followed by the mixer Hamiltonian. (a) QAOA with ZZ-X and DC-QAOA with ZY-X and X-ZY at $p=1$ ($p_1$) and $p=2$ ($p_2$). (b) DC-QAOA with ZY before ZZ term at $p=1$. (c) DC-QAOA with the ZZ before ZY term at $p=1$. AOQ refers to the gate sequence introduced by the AOQMAP approach, while ORIG denotes the original gate sequence. For ZY-X and X-ZY, AOQ-$p_2$ employs the same gate sequence in each depth. All error bars represent one SEM across 10 independent repetitions.}\label{fig:ar_qaoa_dcqaoa}
\end{figure*}
\begin{figure*}
\centering
	\includegraphics[width=\linewidth]{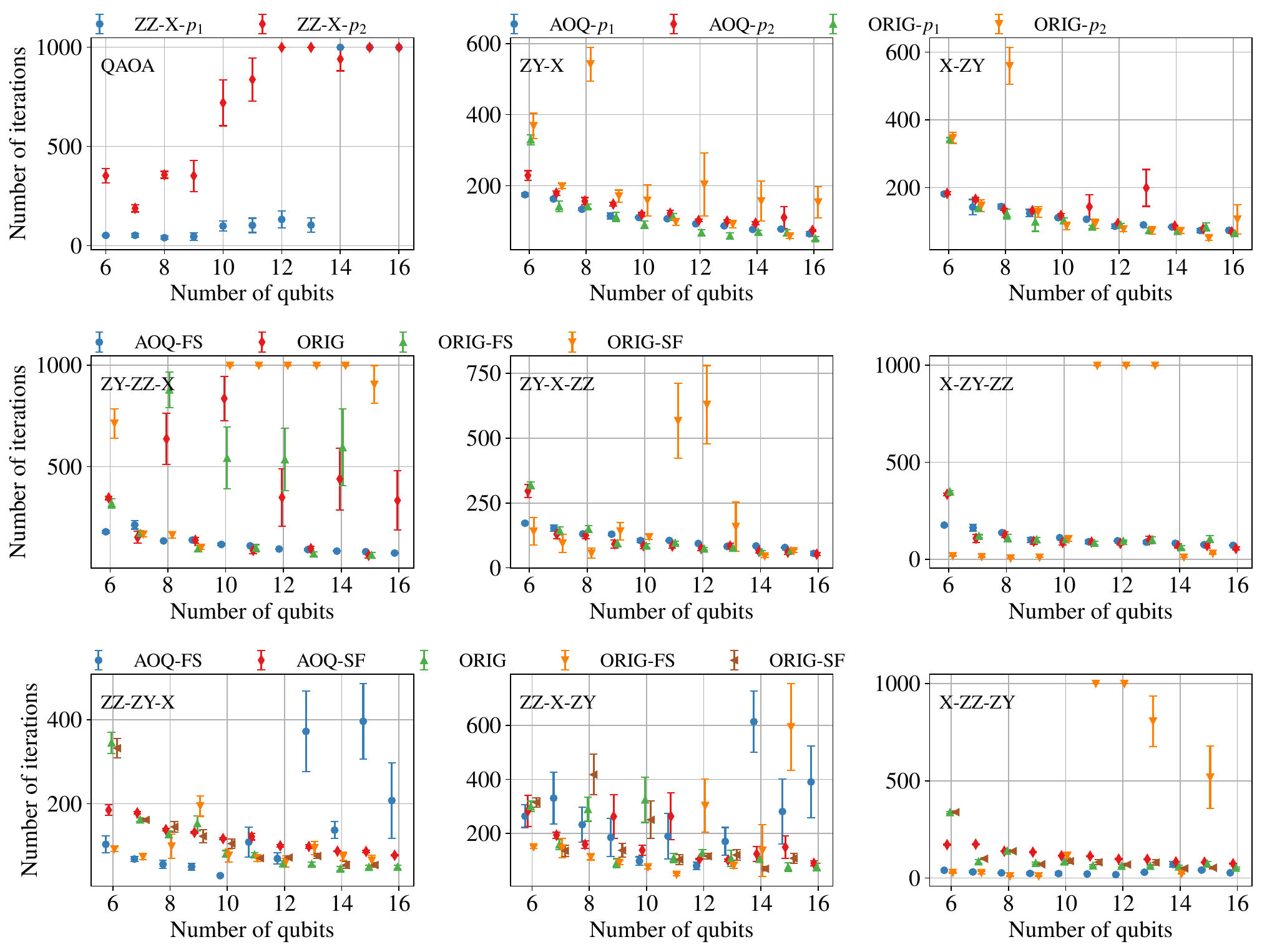}
 \put(-\linewidth,0.73\linewidth){\textbf{(a)}}
 \put(-\linewidth,0.48\linewidth){\textbf{(b)}}
 \put(-\linewidth,0.23\linewidth){\textbf{(c)}}
	\caption{Number of iterations in the optimization process corresponding to Fig.~\ref{fig:ar_qaoa_dcqaoa}. }\label{fig:numb_iter_qaoa_dcqaoa}
\end{figure*}

We investigate the impact of varying Hamiltonian and gate sequences on the performance of DC-QAOA with the ZY CD driving term. Firstly, we examine the effects of applying Hamiltonians in different orders when the problem Hamiltonian is absent. This leads to two possible sequences: ZY-X and X-ZY. Subsequently, we incorporate the problem Hamiltonian, resulting in six distinct Hamiltonian sequences: ZY-ZZ-X, ZY-X-ZZ, X-ZY-ZZ, ZZ-ZY-X, ZZ-X-ZY, and X-ZZ-ZY.

In addition to examining Hamiltonian sequences, we investigate two gate sequences: ORIG and AOQ. For both the first and second two-qubit Hamiltonians, the ORIG sequence employs the original gate sequence, while AOQ utilizes the sequence depicted in block 1 of Fig.~\ref{fig:aoqmap_approach}(e). Within AOQ gate sequence, two specific configurations exist: AOQ-FS and AOQ-SF, illustrated in Figs.\ref{fig:aoqmap_approach}(c) and \ref{fig:aoqmap_approach}(d), respectively. Similarly, within ORIG gate sequence, two configurations are considered: ORIG-FS and ORIG-SF. In ORIG-FS, two-qubit gates in the first two-qubit Hamiltonian term are assigned using ORIG sequence, while the second term employs the reverse order of ORIG sequence. Conversely, in ORIG-SF, the assignment is reversed: the first term utilizes reverse order, while the second term employs ORIG gate sequence. Despite both AOQ-SF and ORIG-SF involving gate sequence reversal, they differ in their approach, with AOQ-SF employing per-layer reversal and ORIG-SF utilizing per-gate reversal.

Figure~\ref{fig:ar_qaoa_dcqaoa} presents the approximation ratio of standard QAOA and DC-QAOA with ZY CD driving term for MaxCut on complete graphs. The number of qubits considered ranges from 6 to 16. For QAOA and the DC-QAOA without problem Hamiltonian, depths are limited to two, while for DC-QAOA including problem Hamiltonian, the depth is set to one. As shown in Fig.~\ref{fig:ar_qaoa_dcqaoa}(a), approximation ratio of QAOA fluctuates with the number of qubits, exhibiting better performance at depth two. The ORIG and AOQ gate sequences in QAOA do not influence the performance since each ZZ gate commutes with the remaining ZZ gates.

For DC-QAOA without problem Hamiltonian, we compare ZY-X and X-ZY with different gate sequences. With ZY applied first (ZY-X), AOQ achieves an overall more stable and improved approximation ratio compared to ORIG. Similarly, AOQ-$p_2$ delivers the highest and most stable approximation ratio for X-ZY.
In AOQ-$p_2$, ZY gates are constructed by repeating the gates from AOQ-$p_1$ for both ZY-X and X-ZY sequences. Implementing this gate sequence requires maintaining swap layers in block 2 of Fig.~\ref{fig:aoqmap_approach}(e) to reverse to the initial qubit order, which is impractical due to noise. However, we can utilize these swap layers in block 2 to construct solutions at depth two, enabling a symmetric implementation of ZY gates, which we discuss later.

Figures~\ref{fig:ar_qaoa_dcqaoa}(b) and \ref{fig:ar_qaoa_dcqaoa}(c) depict the approximation ratio of DC-QAOA when ZY CD driving term is positioned before or after problem Hamiltonian, respectively. It is observed that sequences ZY-ZZ-X with AOQ-FS, ZZ-ZY-X with AOQ-SF, and X-ZZ-ZY with AOQ-SF exhibit better performance. In terms of gate sequence within ZY term, following the first block of Fig.~\ref{fig:aoqmap_approach}(e) yields better results than following the second one. Specifically, when ZY precedes ZZ term (Fig.~\ref{fig:ar_qaoa_dcqaoa}(b)), ORIG-FS gate sequence outperforms ORIG-SF. Similarly, when ZZ term precedes ZY term (Fig.~\ref{fig:ar_qaoa_dcqaoa}(c)), AOQ-SF, ORIG, and ORIG-SF show better performance than AOQ-FS and ORIG-FS. Moreover, AOQ-SF exhibits the most stable and highest performance, while ORIG-SF performs comparably to ORIG.

Figure~\ref{fig:numb_iter_qaoa_dcqaoa} presents the number of iterations required for the optimization process corresponding to results in Fig.~\ref{fig:ar_qaoa_dcqaoa}. As depicted in Fig.~\ref{fig:numb_iter_qaoa_dcqaoa}(a), the number of iterations for QAOA increases rapidly with growing number of qubits, eventually reaching the maximum iteration setting for larger sizes. This trend, coupled with a decrease in approximation ratio, indicates inefficient scaling for QAOA. In contrast, for ZY-X and X-ZY sequences, the number of iterations decreases as qubit number increases, demonstrating better scalability. Furthermore, the majority of iterations remain below 200, providing an advantage over QAOA in terms of computational efficiency.

In DC-QAOA with ZY term applied first (Fig.~\ref{fig:numb_iter_qaoa_dcqaoa}(b)), ZY-ZZ-X with AOQ-FS demonstrates a decreasing number of iterations as qubit count increases, while maintaining a consistently high approximation ratio. Although sequences ZY-X-ZZ and X-ZY-ZZ with AOQ-FS, ORIG, and ORIG-FS exhibit low and stable numbers of iterations, their approximation ratio fluctuates across different qubit counts. This highlights the benefit of utilizing an end located mixer in achieving stable performance across all problem instances. On the other hand, ORIG-SF exhibits a fluctuating iteration, aligning with variations in approximation ratio. In sequences where ZZ term is applied before ZY term (Fig.~\ref{fig:numb_iter_qaoa_dcqaoa}(c)), ZZ-ZY-X generally yields better results compared to ZZ-X-ZY and X-ZZ-ZY, again demonstrating the advantage of employing an end located mixer Hamiltonian. Moreover, ZZ-ZY-X and X-ZZ-ZY, both with AOQ-SF, demonstrate decreasing numbers of iterations as qubit number increases, while maintaining high and stable approximation ratios. Although X-ZZ-ZY with AOQ-FS exhibits low and stable numbers of iterations, the corresponding approximation ratio varies and remains low.

The results highlight the substantial impact of Hamiltonian and gate sequences on algorithm performance. In particular, gate sequence of ZY CD driving term plays a crucial role in achieving high and stable performance in quantum algorithms. Applying the problem Hamiltonian before CD driving term generally leads to higher approximation ratios. In addition, gate sequences introduced by AOQMAP demonstrate an average improved performance compared to original gate sequences.

\subsubsection{Combined and separated implementations}

In the implementation of DC-QAOA with ZY CD driving term including problem Hamiltonian, we can employ a combined gate implementation (see, e.g., \cite{chai2022shortcuts}), where a ZZ gate is followed by a ZY gate, or vice versa. This approach referred to as combined implementation offers the advantage of eliminating the need for the second block of Hamiltonian during the mapping process. Consequently, only block 1 in Fig.~\ref{fig:aoqmap_approach}(e) is required along with $\mathrm{U_1}$ gate, which corresponds to the combination of ZZ and ZY gates. In the following, we evaluate the performance of the combined implementation of ZZ and ZY gates against the separated implementation of ZZ and ZY terms.

\begin{figure}
\centering
	\includegraphics[width=\linewidth]{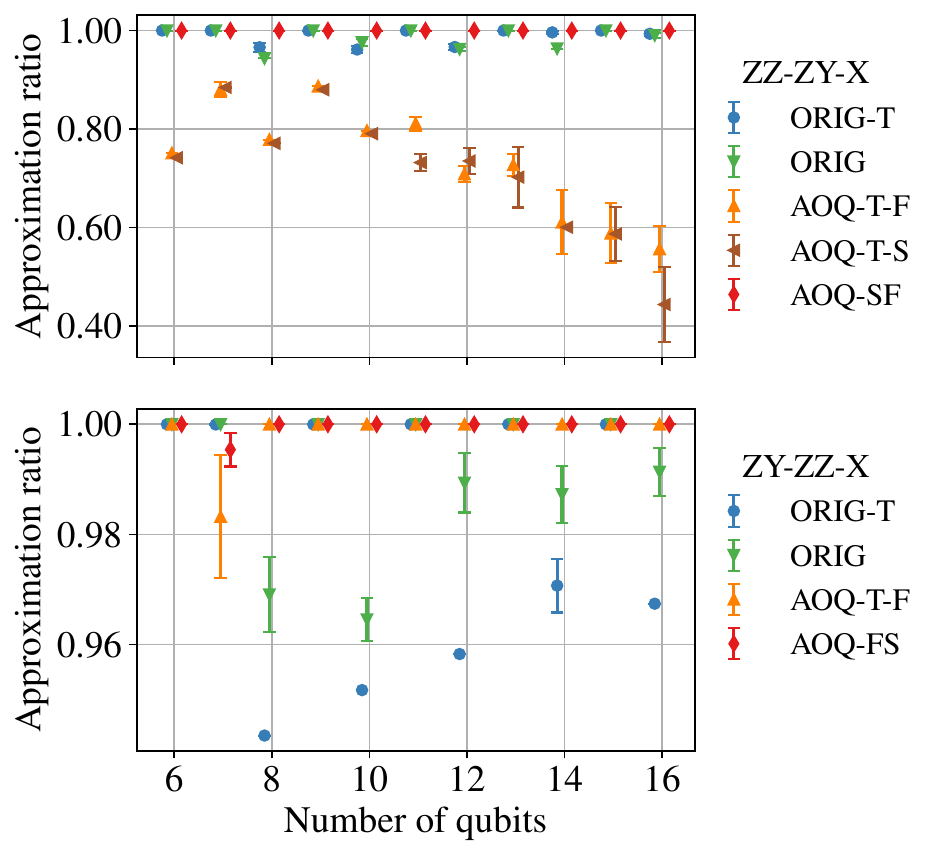}
 \put(-\linewidth,0.888\linewidth){\textbf{(a)}}
 \put(-\linewidth,0.48\linewidth){\textbf{(b)}}
	\caption{Comparison of implementing the ZZ and ZY gates as a combined unit versus separating the implementation with all ZZ gates applied first followed by all ZY gates (or vice versa). ORIG-T denotes the implementation of (a) ZZ-ZY or (b) ZY-ZZ gate as a unit using the original gate sequence, while ORIG represents the separated implementation. AOQ-T-F and AOQ-T-S correspond to the combined implementations using the first and the second block gate sequences from AOQMAP, respectively. AOQ-FS and AOQ-SF refer to the separated implementations. All error bars represent one SEM across 10 independent repetitions.
 }\label{fig:hamiltonian_comb}
\end{figure}
\begin{figure}
\centering
	\includegraphics[width=\linewidth]{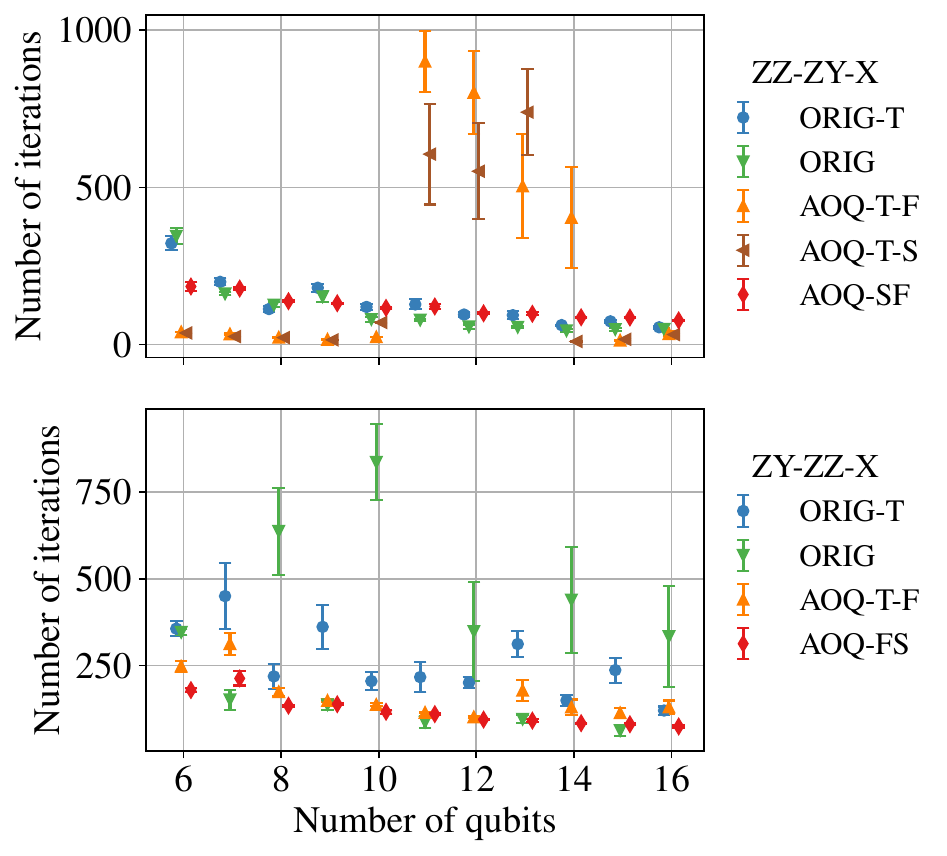}
 \put(-\linewidth,0.885\linewidth){\textbf{(a)}}
 \put(-\linewidth,0.48\linewidth){\textbf{(b)}}
	\caption{Number of iterations in the optimization process corresponding to Fig.~\ref{fig:hamiltonian_comb}. }\label{fig:numb_iter_hamiltonian_comb}
\end{figure}

Figures~\ref{fig:hamiltonian_comb}(a) and \ref{fig:hamiltonian_comb}(b) illustrate approximation ratios of ZZ-ZY-X and ZY-ZZ-X, respectively. For original gate sequence, combined (ORIG-T) and separated (ORIG) implementations of ZZ-ZY-X exhibit comparable performance, while ORIG outperforms ORIG-T for ZY-ZZ-X.
Moreover, for ZZ-ZY-X, separated implementation AOQ-SF surpasses combined implementations AOQ-T-F and AOQ-T-S, which employ gate sequences from block 1 and block 2 of Fig.~\ref{fig:aoqmap_approach}(e), respectively. Similarly, for ZY-ZZ-X, the separated implementation AOQ-FS also outperforms the combined implementation AOQ-T-F.

Figure~\ref{fig:numb_iter_hamiltonian_comb} provides the number of iterations required in the optimization process, corresponding to results presented in Fig.~\ref{fig:hamiltonian_comb}. In Fig.~\ref{fig:numb_iter_hamiltonian_comb}(a), AOQ-T-F and AOQ-T-S, which exhibit lower and fluctuating approximation ratios, demonstrate fluctuations, whereas ORIG-T, ORIG, and AOQ-SF show a more stable and decreasing trend as qubit number increases. For ZY-ZZ-X (Fig.~\ref{fig:numb_iter_hamiltonian_comb}(b)), AOQ-T-F and AOQ-FS, where ZY term follows the sequence in block 1, consistently require a low number of iterations. In contrast, ORIG-T and ORIG produce fluctuating outcomes, aligning with the trends observed in approximation ratios shown in Fig~\ref{fig:hamiltonian_comb}(b).

The results demonstrate that ZZ-ZY-X with AOQ-SF and ZY-ZZ-X with AOQ-FS consistently achieve the highest approximation ratios and require fewer iterations as the number of qubits increases. These findings underscore the advantage of employing separate implementations to enhance the performance of DC-QAOA.

\subsubsection{Symmetry in implementation}

As previously discussed, ZY-X and X-ZY with AOQ gate sequence at depth two (AOQ-$p_2$ in Fig.~\ref{fig:ar_qaoa_dcqaoa}(a)) utilize an identical gate sequence for each depth, necessitating the use of swap layers in block 2 of Fig.~\ref{fig:aoqmap_approach}(e) to restore the initial qubit order. In this study, we investigate alternative implementations of ZY-X and X-ZY at depth two.
Specifically, we examine the effects of symmetric implementations, which hold the potential to improve performance at both the algorithmic level by suppressing Trotter error \cite{tran2021faster,childs2019faster,faehrmann2022randomizing} and the hardware level by enhancing noise resilience in quantum device \cite{ji2023algorithmoriented}. Moreover, symmetry has demonstrated advantages in variational quantum machine learning \cite{meyer2023exploiting} and variational quantum optimization \cite{bravyi2020obstacles, lyu2023symmetry, liu2019variational}.

\begin{figure}
\centering
	\includegraphics[width=\linewidth]{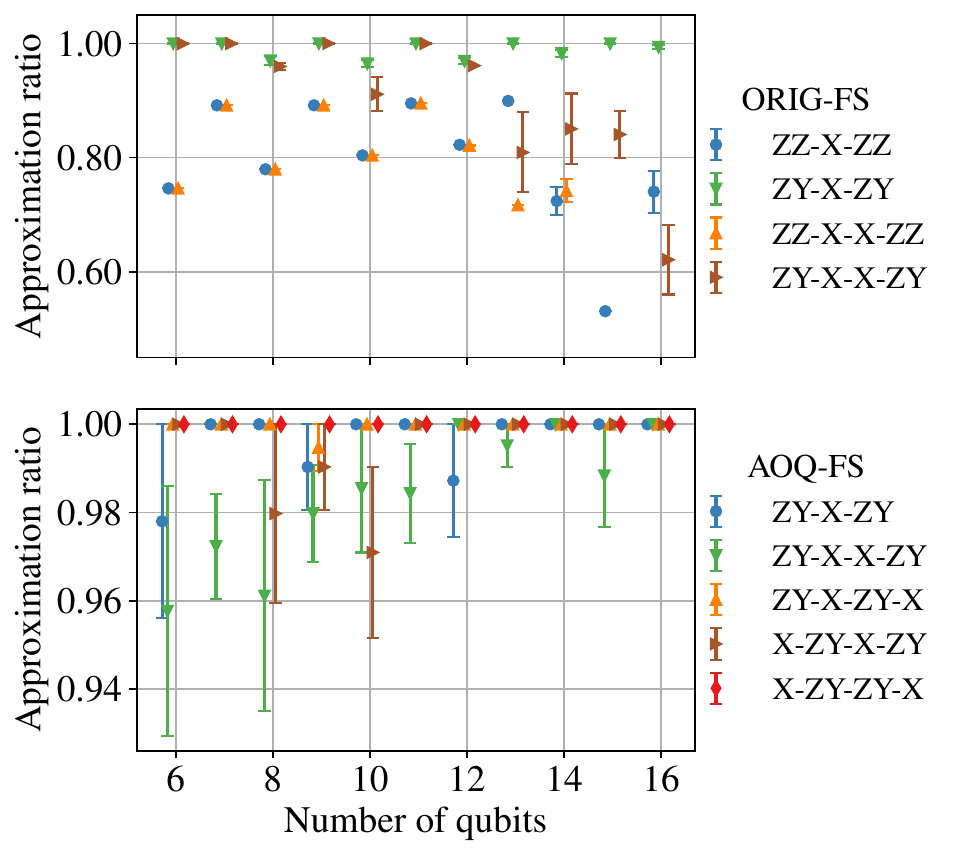}
 \put(-\linewidth,0.88\linewidth){\textbf{(a)}}
 \put(-\linewidth,0.48\linewidth){\textbf{(b)}}
	\caption{Comparison of symmetric implementation using (a) ORIG-FS and (b) AOQ-FS. All error bars represent one SEM across 10 independent repetitions.
 }\label{fig:hamiltonian_symmetry}
\end{figure}
\begin{figure}
\centering
	\includegraphics[width=\linewidth]{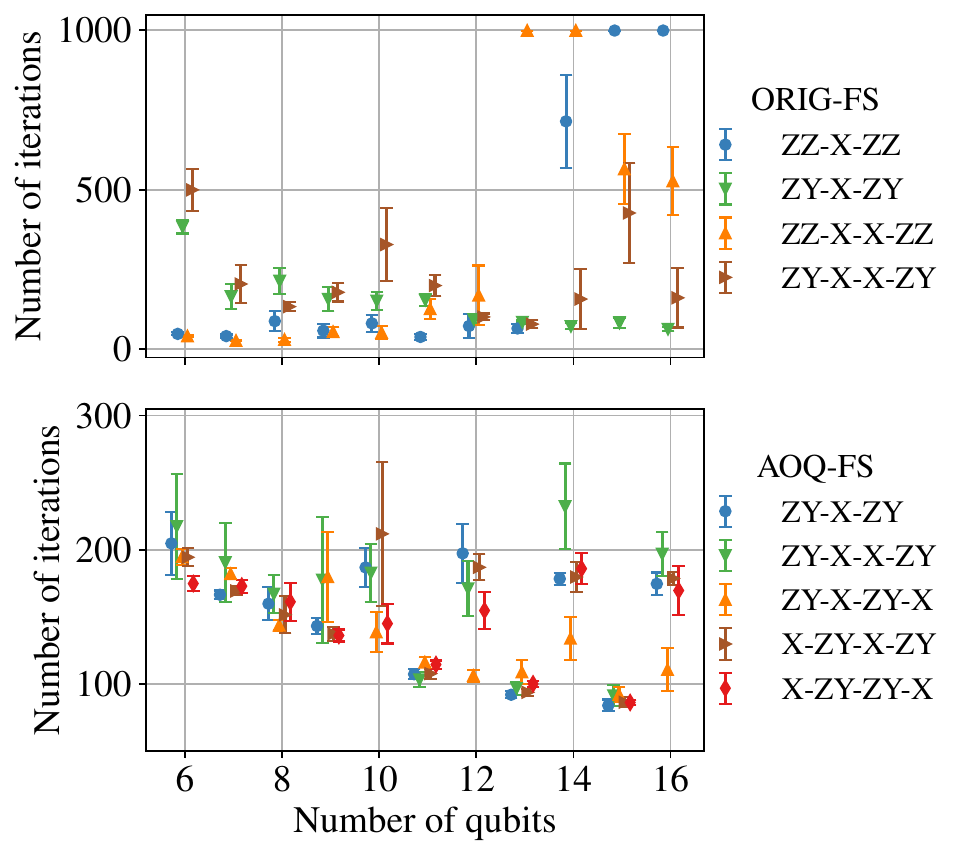}
 \put(-\linewidth,0.88\linewidth){\textbf{(a)}}
 \put(-\linewidth,0.48\linewidth){\textbf{(b)}}
	\caption{Number of iterations in the optimization process corresponding to Fig.~\ref{fig:hamiltonian_symmetry}. }\label{fig:numb_iter_hamiltonian_symmetry}
\end{figure}

Figures~\ref{fig:hamiltonian_symmetry}(a) and \ref{fig:hamiltonian_symmetry}(b) present the approximation ratio achieved using ORIG-FS and AOQ-FS gate sequences, respectively. AOQ-FS generally outperforms ORIG-FS. For ORIG-FS (Fig.~\ref{fig:hamiltonian_symmetry}(a)), ZY-X-ZY performs better than other sequences but fluctuates with the number of qubits. For AOQ-FS (Fig.~\ref{fig:hamiltonian_symmetry}(b)), X-ZY-ZY-X achieves the highest performance, highlighting the effectiveness of the symmetric sequence of ZY gates combined with the mixer Hamiltonian at the beginning and end of the sequence. Additionally, ZY-X-ZY-X demonstrates better performance compared to X-ZY-X-ZY, suggesting that mixer terms located at the end tend to outperform those located at the beginning.

Figure~\ref{fig:numb_iter_hamiltonian_symmetry} shows the number of iterations corresponding to results in Fig.~\ref{fig:hamiltonian_symmetry}. With ORIG-FS (Fig.~\ref{fig:numb_iter_hamiltonian_symmetry}(a)), ZY-X-ZY exhibits a decreasing number of iterations as qubit number increases, consistent with its high performance in Fig.~\ref{fig:hamiltonian_symmetry}(a). In contrast, sequences containing ZZ term demonstrate fluctuating iterations, aligning with their lower approximation ratios. For AOQ-FS (Fig.~\ref{fig:numb_iter_hamiltonian_symmetry}(b)), DC-QAOA with ZY terms requires fewer than 300 iterations. The sequence X-ZY-ZY-X achieves the best performance, requiring fewer than 200 iterations.

These findings underscore the crucial role of the gate sequence in determining the performance of DC-QAOA with symmetric implementation. Strategic placement of mixer terms within the sequence can lead to improved approximation ratios and reduced numbers of iterations.

\subsection{Proposed optimization approach}

\begin{figure}
\centering
	\includegraphics[width=0.7\linewidth]{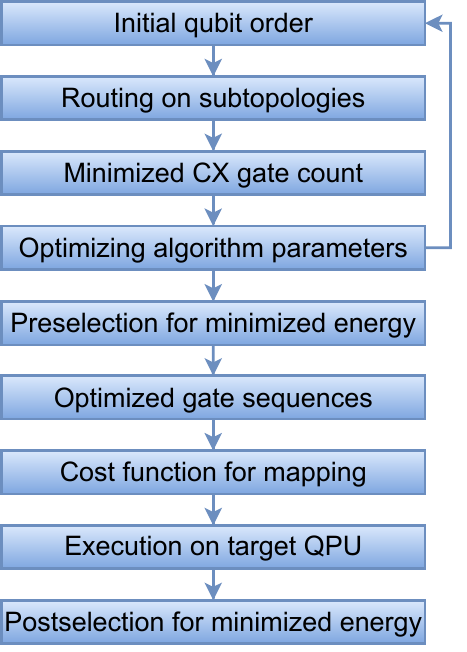}
 \caption{Optimization approach for applying AOQMAP to digitized counterdiabatic quantum optimization algorithms with a deterministic Hamiltonian sequence.
 }\label{fig:optimization_approach}
\end{figure}

The previous results demonstrate that as qubit count increases, a reduction in the number of required iterations can result in stable and high performance with enhanced scalability. Gate sequences generated by AOQMAP provide advantages over original gate sequences in optimizing digitized counterdiabatic quantum algorithms. Moreover, AOQMAP employs a per-layer gate reversal strategy, which outperforms the per-gate reversal strategy used in original gate sequence, as evidenced by enhanced performance of AOQ-FS compared to ORIG-FS for ZY-ZZ-X, and AOQ-SF compared to ORIG-SF for ZZ-ZY-X.
In all implementations, gate sequence of non-all-to-all commuting gates, such as ZY gates, significantly influences the algorithm performance.
Although ZZ-ZY-X with AOQ-SF and ZY-ZZ-X with AOQ-FS achieve comparable high performance, ZY-ZZ-X with AOQ-FS can be obtained with an initialized qubit order $\{0, 1, 2, ..., n-1\}$, whereas ZZ-ZY-X with AOQ-SF requires an optimized initial qubit order (see Fig.~\ref{fig:aoqmap_approach}(e)), which may necessitate more time and effort to determine.

Based on these findings, we propose the method to enhance the performance of digitized counterdiabatic quantum algorithms, which involves cooptimizing algorithm parameters and gate sequences while satisfying hardware connectivity. Figure~\ref{fig:optimization_approach} illustrates {the process for applying AOQMAP to the digitized counterdiabatic quantum algorithms with a deterministic Hamiltonian sequence. We first evaluate routing solutions on subtopologies using different swap layers such as for linear, T-shaped, and H-shaped configurations. Each initial qubit order introduces a unique gate sequence, leading to a fixed number of CX gates. For VQAs with fully connected two-qubit interactions, any arbitrary initial qubit order permits the implementation of all required two-qubit gates. However, when dealing with partially connected two-qubit interactions, optimizing the initial qubit order becomes crucial for reducing CX gates required for the implementation.

After minimizing CX gate count, the next step involves optimizing algorithm parameters using a classical optimizer to minimize the expectation value of the problem Hamiltonian or the energy. A specific number of initial qubit orders, which minimize CX gate count, are evaluated to identify the gate sequence that minimizes the energy. The optimized gate sequences generated from different types of subtopologies may exhibit varying algorithm performances. Therefore, a preselection can be conducted to identify gate sequences from different subtopologies that minimize energy and demonstrate better performance. These selected sequences are then executed on the target QPU with a cost function to assess qubit quality. After the execution, a postselection step can be performed to identify the best mapping scheme from various types of subtopologies by selecting the one that corresponds to the minimum energy.

\section{\label{sec:applic}Applications and evaluation}

We validate the effectiveness of our optimization approach through experimental evaluations conducted on three problem instances: unweighted MaxCut on complete graphs, unweighted MaxCut on noncomplete graphs, and portfolio optimization. The problem Hamiltonian for MaxCut on complete graphs shares similarities with the Hamiltonian representing portfolio optimization. In particular, portfolio optimization corresponds to the weighted MaxCut problem on complete graphs, where edges connecting two nodes have varying weights. This introduces additional complexities compared to the unweighted MaxCut problem.

\subsection{MaxCut on complete graphs}

\begin{figure}
\centering
	\includegraphics[width=0.9\linewidth]{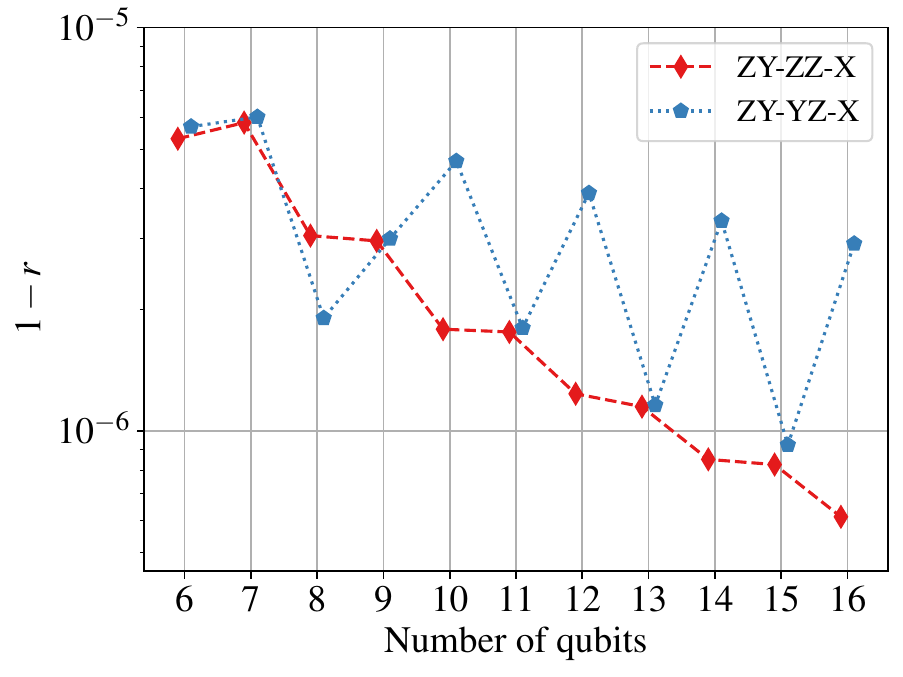}
 \caption{Deviation $1-r$ of approximation ratio of DC-QAOA for MaxCut on complete graphs at $p=1$ with the AOQ-FS gate sequence on a linear topology. A lower deviation indicates a higher approximation ratio, implying improved performance.}\label{fig:maxcut_complete_compare}
\end{figure}

We examine two variations of DC-QAOA for MaxCut on complete graphs. The first variation, ZY-ZZ-X, employs a ZY CD driving term and includes problem Hamiltonian. The second variation, ZY-YZ-X, utilizes a ZY-YZ CD driving term but excludes problem Hamiltonian.
Both sequences follow a consistent Hamiltonian sequence, starting with the CD driving term, followed by problem Hamiltonian, and concluding with the mixer. This specific order provides advantages in dealing with non-all-to-all commuting gates in CD driving terms, which necessitates the optimization of gate sequences.
We set the maximum number of iterations to 200 and employ 10 random initial guesses for the optimization process. Furthermore, we explore 10 initial qubit orders, starting with the default order $\{0, 1, ..., n-1\}$ and generating random permutations for the remaining orders. The order that minimizes the expectation value of the problem Hamiltonian is selected.

Figure~\ref{fig:maxcut_complete_compare} depicts the deviation of the approximation ratio of DC-QAOA for the MaxCut problem on complete graphs, with qubit number ranging from 6 to 16. The gate sequence is according to AOQ-FS on a linear topology. We observe that the deviation for each data point is less than $10^{-5}$, demonstrating a high approximation ratio. Additionally, ZY-ZZ-X sequence exhibits a decreasing deviation as qubit count increases, implying improved performance. In comparison, ZY-YZ-X sequence demonstrates fluctuating results. One possible explanation is that non-all-to-all commuting YZ gates introduce different Trotter errors for odd and even numbered qubits.
Among ten initial qubit orders examined, the default qubit order consistently produces the lowest expectation value for all data points in Fig.~\ref{fig:maxcut_complete_compare}. This finding underscores the robustness and efficacy of the default qubit order in achieving optimal results for this specific problem. Employing a predetermined initial qubit order facilitates the acceleration of the optimization process, as the search for an optimal initial qubit order is eliminated.
Compared to routing solutions obtained using AOQMAP on linear topologies, solutions on T-shaped and H-shaped exhibit a reduced CX gate count but an increased circuit depth due to their enhanced connectivity \cite{ji2023algorithmoriented}. However, identifying an optimal initial qubit order that minimizes energy on these subtopologies may pose a challenge due to their unique characteristics.

\subsection{MaxCut on noncomplete graphs\label{subsec:maxcut_on_noncomp}}

The MaxCut problem on noncomplete graphs is more general and challenging compared to the MaxCut problem on complete graphs. In noncomplete graphs, the absence of edges leads to the lack of corresponding ZZ gates, resulting in reduced symmetry and presenting significant challenges for both classical and quantum optimization algorithms.
Furthermore, in the MaxCut problem on complete graphs, any initial qubit order yields the same additional CX gate count. However, for noncomplete graphs, different initial qubit orders result in varying additional CX gate counts. Consequently, there exists a trade-off between minimizing CX gate count and optimizing gate sequence. As shown in Fig.~\ref{fig:optimization_approach}, our approach starts by searching for an efficient mapping solution that minimizes CX gates on a subtopology. Here, we focus on T-shaped subtopology, but the methodology is also applicable to other subtopologies. We generate a set of $N_o$ random initial qubit orders and select the order that yields the lowest CX gate count. Increasing $N_o$ improves the probability of finding an optimal solution but also poses challenges in handling large problem instances. We set a maximum value of $N_o$ to 3000 for problem instances investigated. This restriction allows us to determine a fixed gate sequence that minimizes CX gates among $N_o$ tests conducted.

\begin{figure}
\centering
	\includegraphics[width=0.9\linewidth]{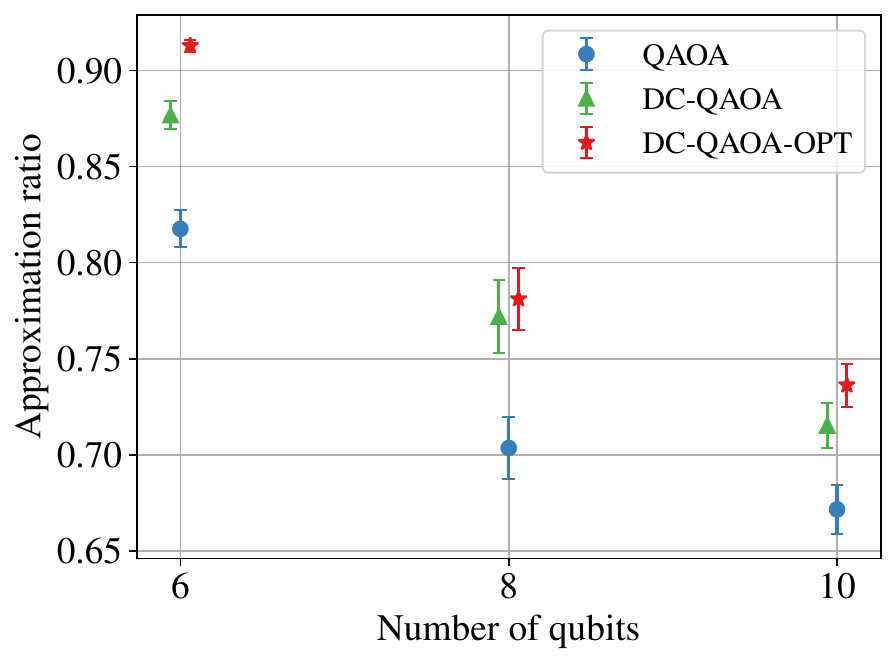}
 \caption{Approximation ratio of QAOA, DC-QAOA, and DC-QAOA-OPT at $p=2$ for MaxCut on 10 randomly generated 3-regular graphs. All error bars represent one SEM.}
 \label{fig:maxcut_noncomplete_ar}
\end{figure}

\begin{figure}
\centering
	\includegraphics[width=0.9\linewidth]{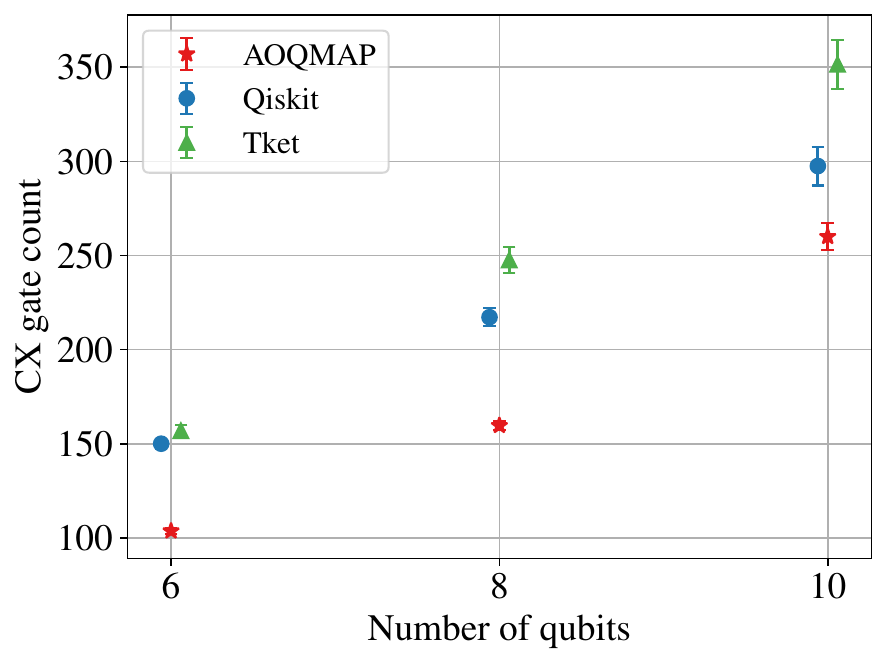}
 \caption{Number of CX gates in the DC-QAOA circuit at $p=2$ for MaxCut on 10 randomly generated 3-regular graphs mapped using AOQMAP, Qiskit, and Tket. All error bars represent one SEM.}
 \label{fig:maxcut_noncomplete_cx}
\end{figure}

QAOA and DC-QAOA use the ORIG gate sequence, while DC-QAOA-OPT uses the AOQ-FS gate sequence. Furthermore, DC-QAOA employs the default Hamiltonian sequence ZZ-X-ZY, whereas DC-QAOA-OPT utilizes the ZY-ZZ-X. All algorithms utilize a gradient descent optimizer with 10 random initial guesses and a maximum number of iterations of 200.
Figure~\ref{fig:maxcut_noncomplete_ar} demonstrates a high performance of DC-QAOA-OPT compared to DC-QAOA and QAOA for the MaxCut problem on three regular graphs. The consistent improvement in the approximation ratio achieved by DC-QAOA-OPT underscores the effectiveness of the proposed optimization approach.
We further highlight the advantages of AOQMAP in reducing CX gate count compared to Qiskit and Tket.
In Qiskit, we use the default setting to compile the circuit directly on a 27-qubit QPU, with the default Hamiltonian and gate sequences. By default, Qiskit employs an optimization level of 1 to compile quantum circuits directly. This default level is chosen to strike a balance between optimization and computational effort. Higher levels may produce higher quality circuits but require more time and computational resources.
In Tket, we use the NoiseAwarePlacement initial mapping and its default routing method \textit{RoutingPass}. Then, we apply Tket's own optimization strategies including \textit{DecomposeBRIDGE}, \textit{DecomposeSWAPtoCX}, and \textit{RemoveRedundancies}. Finally, the default setting of Qiskit's transpiler is used to decompose gates into the backend's basis gate set. Figure~\ref{fig:maxcut_noncomplete_cx} shows comparison results. We observe that AOQMAP consistently produces the fewest number of CX gates compared to Qiskit and Tket. Specifically, AOQMAP achieves an average reduction of 27.2\% in CX gate count. These results highlight the improved performance of AOQMAP at the algorithmic level, while also providing efficient qubit mapping solutions.

\subsection{Portfolio optimization}

\begin{figure*}
\centering
\includegraphics[width=0.85\linewidth]{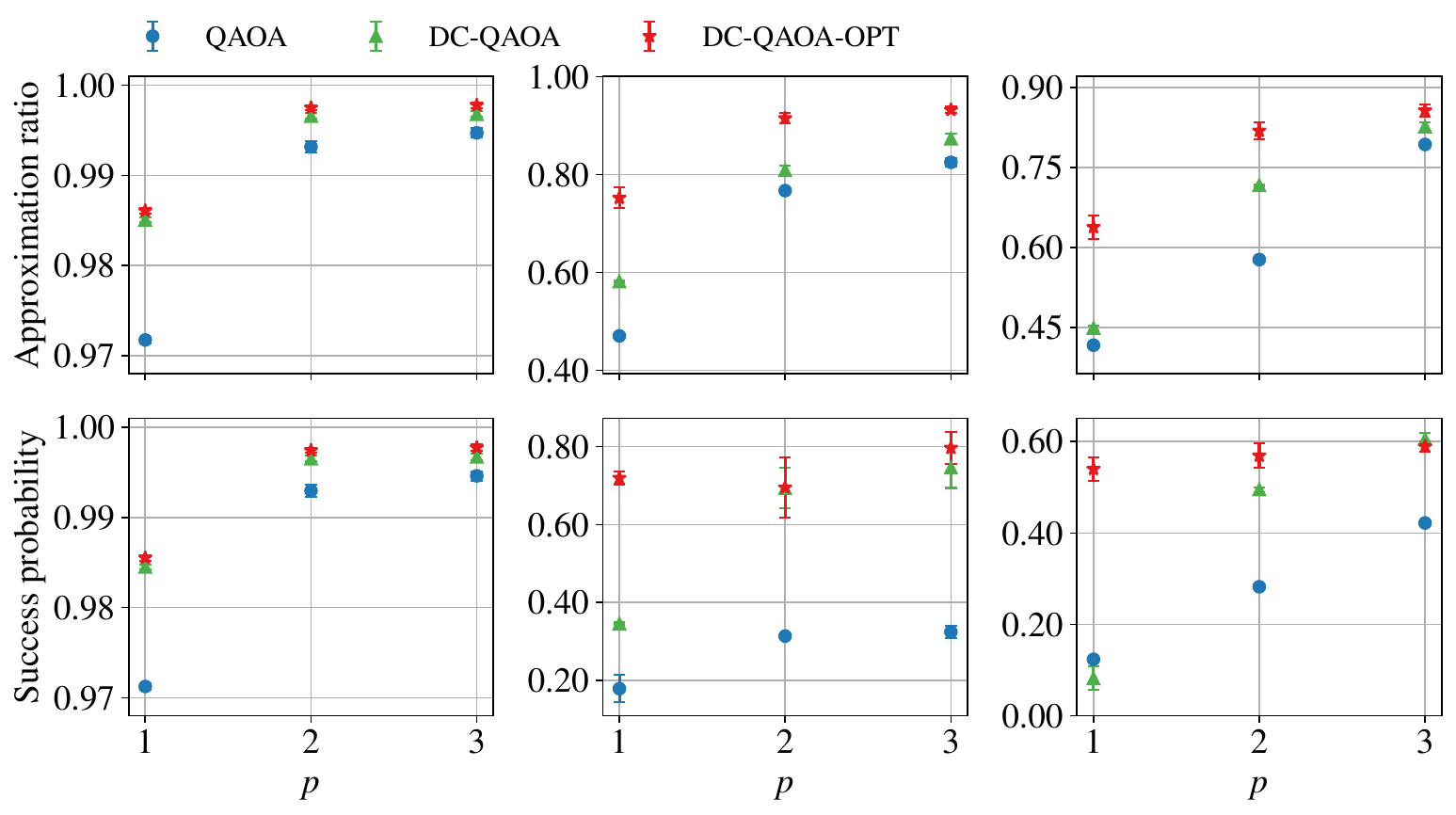}
 \put(-0.85\linewidth,0.44\linewidth){\textbf{(a)}}
 \put(-0.85\linewidth,0.233\linewidth){\textbf{(b)}}
 \caption{(a) Approximation ratio and (b) success probability of QAOA, DC-QAOA, and DC-QAOA-OPT for portfolio optimization with three qubits (left), four qubits (middle), and five qubits (right). DC-QAOA utilizes the ZZ-Z-X-ZY Hamiltonian sequence with the ORIG gate sequence, while DC-QAOA-OPT is optimized over the ZY-ZZ-Z-X, ZZ-Z-X-ZY, and ZY-YZ-Z-X Hamiltonian sequences with AQO-FS gate sequence. All error bars represent one SEM across 10 independent repetitions.
 }\label{fig:portopt_ar_sp}
\end{figure*}

We now investigate the application of QAOA and DC-QAOA in solving portfolio optimization problem, which aims to identify a portfolio that maximizes expected return while minimizing risk, subject to certain constraints \cite{hegade2022portfolio, brandhofer2023benchmarking}.
The portfolio optimization problem involving $n$ assets can be encoded into a problem Hamiltonian, represented by \cite{brandhofer2023benchmarking}
\begin{equation}
H_c = \sum_{i=1}^{n-1}\sum_{j=i+1}^{n} w_{ij} Z_iZ_j + \sum_{i=1}^{n} w_{i} Z_i + w_0,
\label{eq:portopt_prob_hamil}
\end{equation}
where $w_{ij}$, $w_i$, and $w_0$ are coefficients determined by the specific problem instance, $Z_i Z_j$ represents ZZ interaction between qubits $i$ and $j$, and $Z_i$ denotes Pauli $Z$ operator acting on qubit $i$. For both QAOA and DC-QAOA, we employ the standard mixer (Eq.~\ref{eq:mixer}).

The performance of algorithms is assessed using approximation ratio and success probability or ground state probability. Typically, a predetermined number of assets should be chosen, called the budget constraint. The approximation ratio is set to zero if the solution violates this constraint; otherwise, it is defined as
\begin{equation}
r = \frac{E-E_{\max}}{E_0-E_{\max}},
\label{eq:appr_ratio_portopt}
\end{equation}
where $E$ is the expectation value of problem Hamiltonian obtained using QAOA or DC-QAOA, $E_0$ is ground state energy, and $E_{\max}$ signifies the worst-case value. During optimization, the objective is to minimize a predefined expectation value of problem Hamiltonian, considering only those solutions that adhere to budget constraint. If a solution does not satisfy the constraint, its corresponding expectation value is set to zero. Similarly, success probability refers to the probability of obtaining the optimal solution from the set of valid solutions satisfying budget constraint.

We evaluate the performance of QAOA, DC-QAOA, and DC-QAOA-OPT for portfolio optimization instances with a qubit count ranging from 3 to 5 and a depth ranging from 1 to 3. Our primary focus is on assessing the effectiveness of the optimized DC-QAOA utilizing AOQMAP in comparison to conventional DC-QAOA and standard QAOA approach.
The default DC-QAOA utilizes ZZ-Z-X-ZY Hamiltonian sequence with ORIG gate sequence, repeatedly applied $p$ times to reach a depth of $p$. In contrast, DC-QAOA-OPT explores three Hamiltonian sequences with AOQ-FS gate sequence: ZY-ZZ-Z-X, ZZ-Z-X-ZY, and ZY-YZ-Z-X. The sequence minimizing energy is selected. For ZY-ZZ-Z-X, ZY CD driving term is applied first, followed by problem Hamiltonian, and finally the mixer Hamiltonian. In ZZ-Z-X-ZY, ZY CD driving term is placed at the end. In comparison, ZY-YZ-Z-X employs a ZY-YZ CD driving term while lacking the two-qubit Hamiltonian in problem Hamiltonian. In both DC-QAOA and DC-QAOA-OPT, CD driving term follows the same coefficient as the ZZ term in problem Hamiltonian (Eq.~\ref{eq:portopt_prob_hamil}).

The COBYLA algorithm is employed in the optimization process, with a maximum number of function evaluations (maxiter) set to 1000. The optimization strategy for QAOA follows the approach outlined in Ref.~\cite{brandhofer2023benchmarking}. For DC-QAOA and DC-QAOA-OPT, the strategy for the QAOA part (ZZ-Z-X) is maintained, while randomness is introduced into the CD driving term. To further enhance the exploration of the solution space, $N_I$ randomly generated initial guess values are additionally incorporated into the optimization process for QAOA, DC-QAOA, and DC-QAOA-OPT. In our case, $N_I$ is set to 30.
Similar to the MaxCut on complete graphs, any initial qubit order allows for the implementation of all two-qubit gates in DC-QAOA for portfolio optimization but results in different gate sequences, leading to variations in optimization efficiency. DC-QAOA-OPT employs the optimization process shown in Fig.~\ref{fig:optimization_approach} to generate an optimized gate sequence. This involves evaluating $N_o$ initial qubit orders to identify the one that minimizes the expectation value of problem Hamiltonian. In this study, ten different initial qubit orders are utilized to optimize the gate sequence on a linear topology.
The optimization process is repeated ten times, and results are presented with error bars that represent one SEM.

Figure~\ref{fig:portopt_ar_sp} depicts the approximation ratio and success probability of QAOA, DC-QAOA, and DC-QAOA-OPT for portfolio optimization with varying numbers of qubits. The left, middle, and right subfigures correspond to instances with three, four, and five qubits, respectively. The results clearly demonstrate that DC-QAOA-OPT outperforms DC-QAOA and QAOA in terms of both approximation ratio and success probability, highlighting the advantages of cooptimizing algorithm parameters, gate sequences, and qubit mapping.

To summarize, we propose a methodology for efficiently applying the AOQMAP to DC-QAOA, leveraging the Suzuki-Trotter decomposition to optimize qubit mapping and gate sequences. The AOQMAP approach introduces a customized gate sequence tailored to the initial qubit order and Hamiltonian sequence. Our approach generates optimized gate sequences that minimize SWAP gates, improving the resilience of algorithms on noisy quantum devices. Furthermore, these optimized gate sequences facilitate efficient optimization of algorithm parameters, leading to improved performance at the algorithmic level compared to the default sequence. Extensive experiments demonstrate that optimizing sequences of Hamiltonian and non-all-to-all commuting gates leads to substantial performance gains, as evidenced by increased approximation ratio and reduced iterations as the qubit count increases. The effectiveness of our approach is further confirmed by its successful application to other problem instances, showcasing its ability to enhance performance and pave the way for further advancements in DC-QAOA.

\subsection{Qubit mapping strategy evaluation}

\begin{figure*}
\centering
	\includegraphics[width=\linewidth]{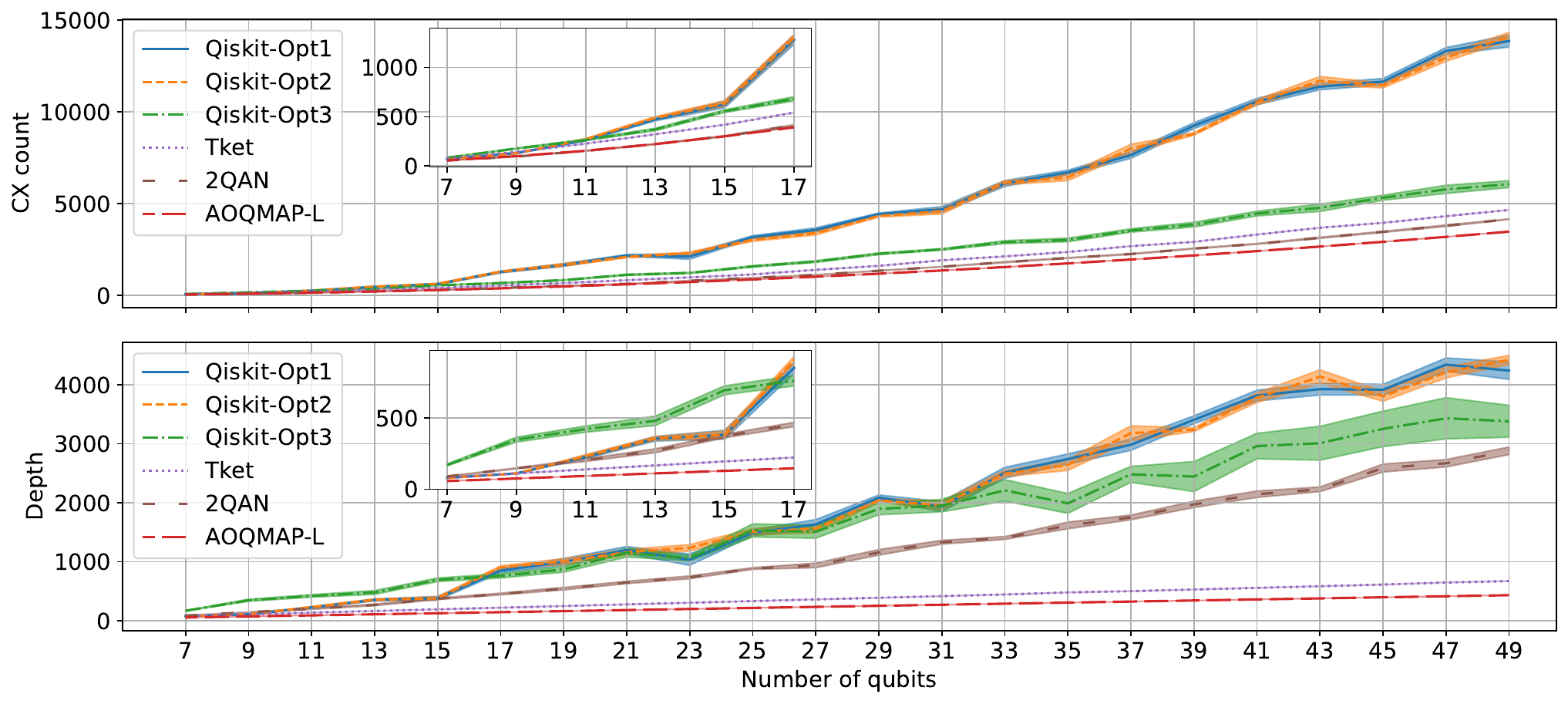}
 \caption{Number of CX gates and circuit depth of mapped QAOA for MaxCut on complete graphs with $p=1$ and qubit counts ranging from 6 to 50 using Qiskit with optimization levels 1 (Opt1), 2 (Opt2), and 3 (Opt3), Tket, 2QAN, and AOQMAP-L. Lower values mean better results.}\label{fig:circ_properties}
\end{figure*}

\begin{table*}
\caption{\label{tab:circ_properties}%
Average (avg) and maximum (max) reduction in CX gates and circuit depth using AOQMAP-L compared to other qubit mapping strategies in Fig.~\ref{fig:circ_properties}.
}
\begin{ruledtabular}
\begin{tabular}{lcccccccccc}
& \multicolumn{2}{c}{Qiskit-Opt1} & \multicolumn{2}{c}{Qiskit-Opt2} & \multicolumn{2}{c}{Qiskit-Opt3} & \multicolumn{2}{c}{Tket} & \multicolumn{2}{c}{2QAN} \\
& avg & max & avg & max & avg & max & avg & max & avg & max\\
\colrule
CX count & 65.1\% & 77.1\% & 65.2\% & 77.2\% & 43.4\% & 47.7\% & 27.3\% & 31.7\% & 8.8\% & 16.3\%\\
Depth & 79.0\% & 90.5\% & 79.2\% & 90.8\% & 83.3\% & 87.9\% & 34.4\% & 35.9\% & 72.6\% & 84.9\%\\
\end{tabular}
\end{ruledtabular}
\end{table*}

In addition to demonstrating performance improvement of our optimized gate sequence over the default one, we compare our optimization approach to commonly used qubit mapping strategies, including industrial compilers Qiskit \cite{qiskit} and Tket \cite{sivarajah2020tket}, as well as the application-specific compiler 2QAN \cite{lao20222qan}. Effectiveness is evaluated by measuring circuit depth and CX gate count, with lower values indicating higher efficiency.
The 2QAN method leverages the flexibility of permuting gates within Hamiltonian to find solutions that minimize SWAP gates \cite{lao20222qan}. It compiles only the first Trotter step and utilizes the same circuit for odd steps while reversing gate order for even steps, exhibiting a similarity to AOQMAP's utilization of mirror effects. This suggests that for comparison of 2QAN and AOQMAP with respect to circuit properties (excluding optimized gate sequences), evaluating depth-one QAOA circuits suffices. However, unlike our approach that also optimizes gate sequences, 2QAN solely minimizes SWAP gates. For Qiskit and Tket, we apply our optimized sequences to original QAOA circuits for fairness. Again, depth-one QAOA circuits remain sufficient for evaluation. This is because, for depth-two QAOA, Qiskit and Tket typically require more than double the number of SWAP compared to depth one, while 2QAN and AOQMAP guarantee double SWAP gates for depth two.
Finally, focusing solely on QAOA is sufficient since any solution found for QAOA can be extended to DC-QAOA using the strategy proposed in Fig.~\ref{fig:aoqmap_approach}.

The QAOA for MaxCut on complete graphs are first employed as benchmarks. In this case, AOQMAP provides direct solutions for these all-to-all connected two-qubit gates. We map QAOA circuits with a depth of one onto IBM's 127-qubit QPU without considering device's noise characteristics. The QPU's topology can be found in Ref.~\cite{ji2023algorithmoriented}. Its native gate set includes \{ECR, ID, RZ, SX, X\}, where ID is identity gate, RZ is single-qubit rotation gate around $z$-axis, X is Pauli $X$ gate, and SX is square root of X gate. However, here we decompose QAOA into gate set \{CX, RZ, RX, RY\}, where RX and RY are single rotation gates around $x$-axis and $y$-axis, respectively. This choice prioritizes the mapping process by minimizing influence of decomposition and optimization introduced by Qiskit's transpiler. For comparison, we consider Qiskit with optimization levels from 1 to 3. For Tket, we utilize GraphPlacement for initial qubit mapping, followed by its default RoutingPass for routing. Its subsequent optimization strategies are described in Sec.~\ref{subsec:maxcut_on_noncomp}. For 2QAN, the initial mapping strategy employs Qiskit's SABRE mapper due to the computational inefficiency of QAP within 2QAN framework for circuits with larger numbers of qubits and QPUs with many qubits, followed by 2QAN's own routing strategy. In AOQMAP, we employ the solutions on linear (L) subtopologies. T- and H-shaped sub-topologies offer increased connectivity, reducing the number of SWAP gates required while increasing the depth of mapped circuits. Qiskit's transpiler with an optimization level of 1 performs the final decomposition and optimization for Tket, 2QAN, and AOQMAP.

Figure~\ref{fig:circ_properties} presents the number of CX gates and circuit depth of mapped QAOA for qubit numbers ranging from 7 to 49 with various mapping strategies. Within Qiskit, optimization levels 1 (Qiskit-Opt1) and 2 (Qiskit-Opt2) show minimal differences and produce the most CX gates and deepest circuit depths, indicating the lowest solution quality. While Qiskit-Opt3 offers improvement, Tket demonstrates lower CX counts and shallower circuit depths than Qiskit-Opt3. In comparison, 2QAN achieves fewer CX gates but deeper circuit depths than Tket. Finally, AOQMAP-L demonstrates the highest performance with the fewest CX gates and shallowest circuit depths.
Table~\ref{tab:circ_properties} summarizes the reduction in CX count and circuit depth achieved by AOQMAP-L compared to other methods. On average, AOQMAP-L reduces CX gates by 65\% (up to 77\%) compared to Qiskit-Opt1 and Qiskit-Opt2. Similarly, the circuit depth reduction is substantial, averaging 79\% with a maximum of 91\%. Compared to Qiskit-Opt3, AOQMAP-L achieves an average reduction of 43\% in CX count and 83\% in circuit depth. Furthermore, AOQMAP-L outperforms Tket with an average 27\% reduction in the number of CXs and 34\% reduction in circuit depth. Finally, compared to 2QAN, AOQMAP-L shows an average reduction of 9\% in CX count and 73\% in circuit depth, with maximum reductions of 16\% and 85\%, respectively. Overall, AOQMAP-L demonstrates a significant average reduction of 42\% in CX count and 70\% in circuit depth compared to others.

\begin{figure}[htb]
\centering
	\includegraphics[width=\linewidth]{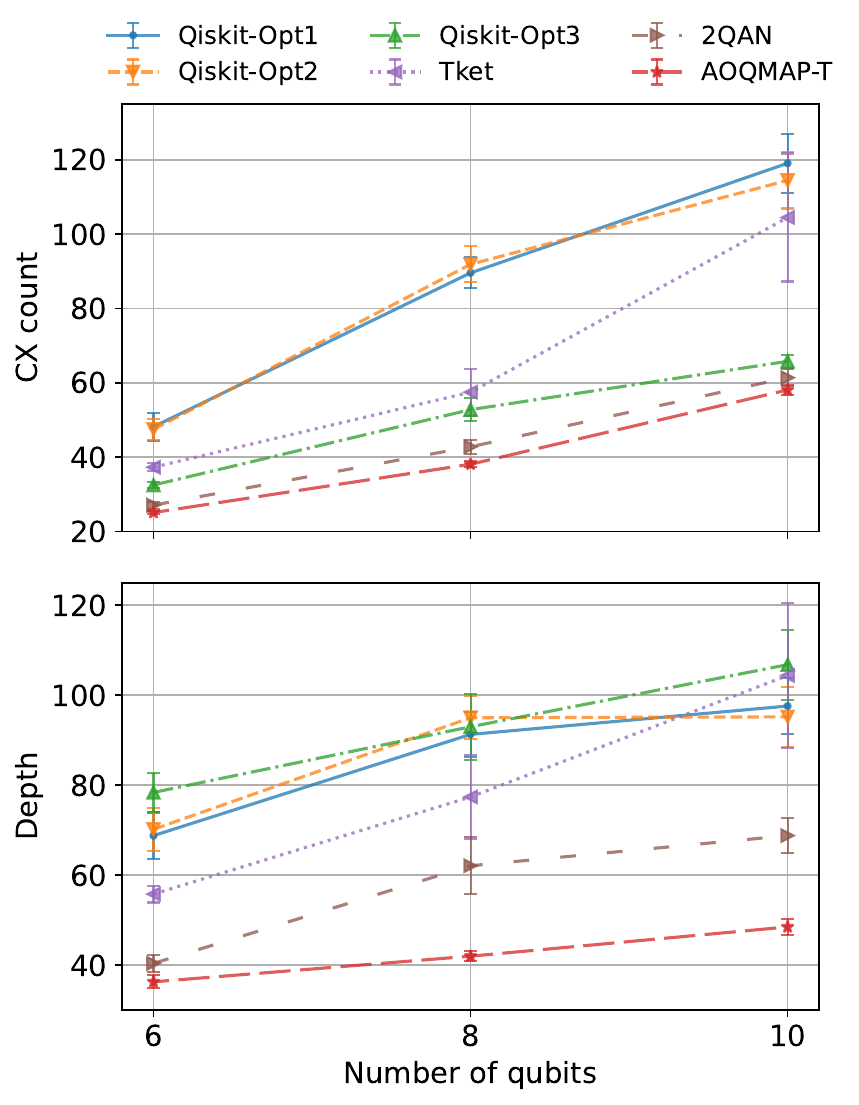}
 \caption{Number of CX gates and circuit depth of mapped QAOA for MaxCut on 
10 randomly generated 3-regular graphs with $p=1$ and qubit counts ranging from 6 to 10 using Qiskit with optimization levels 1 (Opt1), 2 (Opt2), and 3 (Opt3), Tket, 2QAN, and AOQMAP-L. Lower values mean better results. The results are obtained with the maximum numbers of initial qubit orders set to 3000, 20000, and 30000 for 6-qubit, 8-qubit, and 10-qubit QAOA.}\label{fig:circ_properties_noncomplete}
\end{figure}

\begin{table*}[htb]
\caption{\label{tab:circ_properties_noncomplete}%
Average (avg) and maximum (max) reduction in CX gates and circuit depth using AOQMAP-T compared to other qubit mapping strategies in Fig.~\ref{fig:circ_properties_noncomplete}.
}
\begin{ruledtabular}
\begin{tabular}{lcccccccccc}
& \multicolumn{2}{c}{Qiskit-Opt1} & \multicolumn{2}{c}{Qiskit-Opt2} & \multicolumn{2}{c}{Qiskit-Opt3} & \multicolumn{2}{c}{Tket} & \multicolumn{2}{c}{2QAN} \\
& avg & max & avg & max & avg & max & avg & max & avg & max\\
\colrule
CX count & 52.2\% & 57.5\% & 51.7\% & 58.5\% & 20.8\% & 27.8\% & 36.9\% & 44.4\% & 7.9\% & 11.0\%\\
Depth & 50.5\% & 54.0\% & 51.0\% & 55.8\% & 54.3\% & 54.8\% & 44.7\% & 53.5\% & 24.0\% & 32.4\%\\
\end{tabular}
\end{ruledtabular}
\end{table*}

We now evaluate the performance of QAOA for MaxCut problem on 3-regular graphs. In this study, we employ AOQMAP on T-shaped topologies (AOQMAP-T) to generate mapped circuits. The strategy involves optimizing the initial qubit order to minimize the number of CX gates.
It's important to note that for an $n$-qubit QAOA, there are $n!/2$ distinct initial qubit orders due to the inherent symmetry, which is challenging, particularly for larger qubit systems. Therefore, it is important to develop efficient heuristic algorithms for optimizing this task, which is a promising area for future research.

To ensure a fair comparison across different strategies, all methods (Qiskit, Tket, and 2QAN) utilize identical, optimized gate sequences generated by AOQMAP-T. As depicted in Fig.~\ref{fig:circ_properties_noncomplete}, AOQMAP-T generates circuits with the fewest CX gates and the shallowest depth, followed by 2QAN. While Qiskit with optimization level 3 (Qiskit-Opt3) achieves a lower CX count than Tket, it produces circuits with the deepest depth. Optimization levels 1 and 2 in Qiskit exhibit similar performance but result in the highest CX count. Notably, AOQMAP-T achieves an average maximum reduction of 52.2\% in CX count compared to Qiskit-Opt1 and an average maximum decrease of 54.3\% in circuit depth compared to Qiskit-Opt3, as detailed in Table~\ref{tab:circ_properties_noncomplete}.

\section{\label{sec:bench}Benchmarking experiments}

In this section, we assess the efficiency of various qubit mapping techniques for DC-QAOA applied to MaxCut on complete graphs and portfolio optimization. We conduct a comprehensive benchmarking of their performance across a range of metrics, analyze the influence of noise, evaluate the effectiveness of error mitigation strategies, and demonstrate their implementation on IBM QPUs, providing valuable insights into the practical applicability of these mapping strategies.

\subsection{Experimental setup and evaluation metrics}

For AOQMAP, we employ the optimization strategy developed in Sec.~\ref{sec:method} to determine the optimal parameters and gate sequence on a linear subtopology.
For Qiskit, we utilize the default setting (optimization level 1).
For Tket, we use the same settings as in Sec.~\ref{sec:applic}.
While Qiskit and Tket directly map circuit onto target QPU, AOQMAP requires selecting high-quality qubits for execution, which is performed in two stages \cite{ji2023algorithmoriented}. First, mapomatic \cite{nation2023suppressing} is employed to identify all subtopologies that satisfy connectivity constraints of the resulting circuit. Then, a cost function is used to select the subtopology that minimizes error by calculating circuit's error on each subtopology \cite{ji2023algorithmoriented}. For a fair evaluation, all mapped circuits are decomposed onto hardware native basis gate set and optimized using Qiskit transpiler with the default setting.
The properties of QPUs can be found in Appendix~\ref{app:cloud_platf}. It is worth noting that as late as possible (ALAP) scheduling method, customizable in Qiskit's transpiler, is consistently applied across all experiments.

Several metrics are employed to assess the performance of different qubit mapping strategies. The number of CX gates and circuit depth in the resulting circuit are used to compare the efficiency of qubit mapping approaches. Additionally, the approximation ratio is used to evaluate the algorithm performance. A lower CX gate count and shorter circuit depth indicate a more efficient mapping approach, while a higher approximation ratio suggests improved performance. In the absence of noise, Qiskit and Tket should produce similar results as they map the same original circuit. Conversely, AOQMAP generates optimized gate sequences that differ from the original one. In the presence of noise, the performance with different qubit mapping strategies varies due to the distinct number of CX gates in the resulting circuits, reflecting the quality of the mapping process.

\subsection{Circuit properties\label{subsec:circu_proper}}

\begin{table*}[htp]
\caption{\label{tab:qubit_mapping_maxcut_complete}%
Circuit properties of $n$-qubit DC-QAOA at $p=1$ for MaxCut on complete graphs with AOQMAP, Qiskit, and Tket approaches on a 7-qubit QPU ibm\_perth.
}
\begin{ruledtabular}
\begin{tabular}{lcccccc}
&
\multicolumn{2}{c}{\textrm{AOQMAP}}&
\multicolumn{2}{c}{\textrm{Qiskit}}&
\multicolumn{2}{c}{\textrm{Tket}}\\
$n$ & CX count & depth & CX count & depth & CX count & depth \\
\colrule
3 & 14 & 35 & 18 & 44 & 18 & 44\\
4 & 30 & 46 & 46 & 68 & 33 & 73\\
5 & 52 & 59 & 73 & 116 & 69 & 96\\
\end{tabular}
\end{ruledtabular}
\caption{\label{tab:qubit_mapping_maxcut_complete_ehningen}%
Circuit properties on a 27-qubit QPU ibmq\_ehningen. Same as in Table \ref{tab:qubit_mapping_maxcut_complete}.
}
\begin{ruledtabular}
\begin{tabular}{lcccccc}
&
\multicolumn{2}{c}{\textrm{AOQMAP}}&
\multicolumn{2}{c}{\textrm{Qiskit}}&
\multicolumn{2}{c}{\textrm{Tket}}\\
$n$ & CX count & depth & CX count & depth & CX count & depth \\
\colrule
3 & 14 & 35 & 18 & 44 & 18 & 44\\
4 & 30 & 47 & 33 & 73 & 33 & 73\\
5 & 52 & 56 & 73 & 116 & 72 & 96\\
\end{tabular}
\end{ruledtabular}
\end{table*}
\begin{table*}[htp]
\caption{\label{tab:qubit_mapping_portopt}%
Circuit properties of $n$-qubit DC-QAOA with a depth ranging from 1 to 3 for portfolio optimization using AOQMAP, Qiskit, and Tket approaches on a 27-qubit QPU ibm\_cairo.
}
\begin{ruledtabular}
\begin{tabular}{lcccccc}
&
\multicolumn{2}{c}{\textrm{AOQMAP}}&
\multicolumn{2}{c}{\textrm{Qiskit}}&
\multicolumn{2}{c}{\textrm{Tket}}\\
$n$ & CX count & depth & CX count & depth & CX count & depth \\
\colrule
3 & (14, 28, 42) & (32, 60, 88) & (18, 39, 60) & (42, 83, 124) & (18, 39, 60) & (42, 83, 124)\\
4 & (30, 60, 90) & (47, 85, 144) & (42, 84, 126) & (75, 150, 221) & (42, 84, 126) & (75, 142, 208)\\
5 & (52, 104, 156) & (67, 108, 180) & (76, 153, 237) & (107, 199, 322) & (71, 149, 224) & (82, 175, 265)\\
\end{tabular}
\end{ruledtabular}
\caption{\label{tab:qubit_mapping_portopt_ehningen}%
Circuit properties on a 27-qubit QPU ibmq\_ehningen. Same as in Table~\ref{tab:qubit_mapping_portopt}.
}
\begin{ruledtabular}
\begin{tabular}{lcccccc}
&
\multicolumn{2}{c}{\textrm{AOQMAP}}&
\multicolumn{2}{c}{\textrm{Qiskit}}&
\multicolumn{2}{c}{\textrm{Tket}}\\
$n$ & CX count & depth & CX count & depth & CX count & depth \\
\colrule
3 & (14, 28, 42) & (32, 60, 88) & (18, 39, 60) & (42, 83, 124) & (18, 39, 60) & (42, 83, 124)\\
4 & (30, 60, 90) & (47, 89, 138) & (42, 91, 126) & (75, 131, 221) & (42, 84, 126) & (77, 142, 212)\\
5 & (52, 104, 156) & (66, 108, 180) & (73, 152, 228) & (94, 205, 301) & (73, 148, 223) & (94, 184, 274)\\
\end{tabular}
\end{ruledtabular}
\end{table*}

We now investigate circuit properties of DC-QAOA with AOQMAP and compare them to properties obtained using Qiskit and Tket with the experimental setup described previously. Table~\ref{tab:qubit_mapping_maxcut_complete} presents the numbers of CX gates and circuit depths in the final circuit of DC-QAOA with qubit count ranging from 1 to 3 for MaxCut on complete graphs using AOQMAP, Qiskit, and Tket on a 7-qubit QPU ibm\_perth. We observe that AOQMAP consistently produces the fewest number of CX gates and the shortest circuit depth compared to Qiskit and Tket. Specifically, AOQMAP reduces CX gate count by an average of 28.6\% and 18.7\%, and circuit depth by 34\% and 32\% compared to Qiskit and Tket, respectively. These results highlight the effectiveness of AOQMAP in optimizing the circuit properties, resulting in more efficient implementations of algorithms on the QPU.

Table~\ref{tab:qubit_mapping_maxcut_complete_ehningen} presents results for the same problem instances as Table~\ref{tab:qubit_mapping_maxcut_complete} on a 27-qubit QPU ibmq\_ehningen. While CX gate count remains unchanged with AOQMAP, circuit depth varies slightly between different QPUs. In contrast, Qiskit and Tket result in different CX gate counts and circuit depths on two devices. Furthermore, AOQMAP achieves an average reduction of 20\% and 19.7\% in CX gate count, and an average reduction of 36\% and 32.6\% in circuit depth compared to Qiskit and Tket, respectively. These findings demonstrate the consistent ability of AOQMAP to enhance circuit properties across different QPUs.

Tables~\ref{tab:qubit_mapping_portopt} and \ref{tab:qubit_mapping_portopt_ehningen} present circuit properties of DC-QAOA for portfolio optimization on two 27-qubit QPUs ibm\_cairo and ibmq\_ehningen, respectively. Similarly, AOQMAP consistently achieves the lowest CX gate count and the shortest circuit depth compared to Qiskit and Tket.
Specifically, on ibm\_cairo, AOQMAP reduces CX gate count by 29.3\% and 18.2\%, and circuit depth by 36\% and 30\% compared to Qiskit and Tket, respectively. On ibmq\_ehningen, AOQMAP achieves an average reduction of 29.3\% and 28.2\% in CX gate count compared to Qiskit and Tket, respectively, and 33.9\% and 33.2\% in circuit depth.

In summary, the AOQMAP approach yields an average reduction of 28.8\% in the CX gate count and 33.4\% in circuit depth across all problem instances when compared to Qiskit and Tket on the tested QPUs, demonstrating its potential to enhance the performance of DC-QAOA in various applications.

\begin{figure*}[ht]
\centering
	\includegraphics[width=\linewidth]{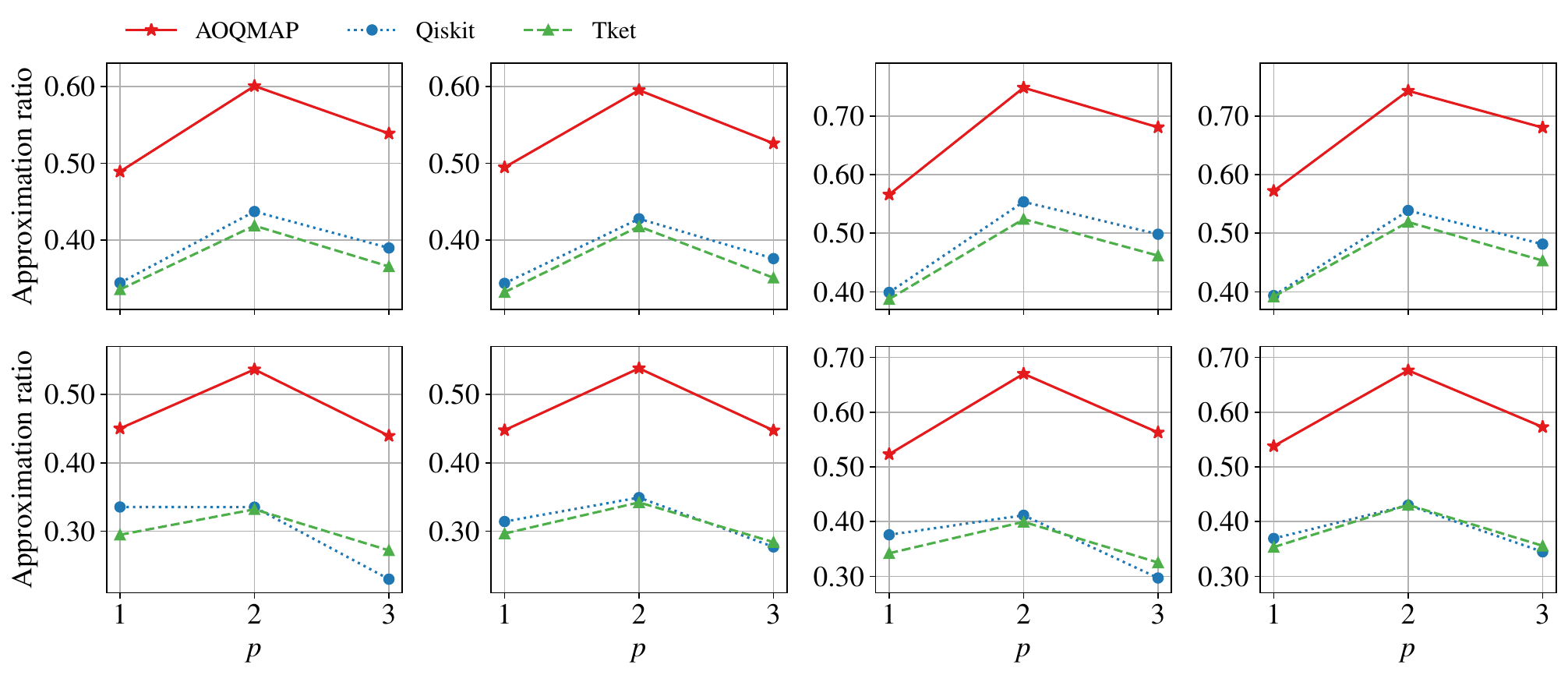}
 \put(-\linewidth,0.4\linewidth){\textbf{(a)}}
 \put(-\linewidth,0.216\linewidth){\textbf{(b)}}
 \put(-0.81\linewidth,0.24\linewidth){\footnotesize Default}
 \put(-0.535\linewidth,0.24\linewidth){\footnotesize DD}
 \put(-0.30\linewidth,0.24\linewidth){\footnotesize ZNE}
 \put(-0.09\linewidth,0.24\linewidth){\footnotesize DD+ZNE}
 \put(-0.823\linewidth,29){\footnotesize Default}
 \put(-0.535\linewidth,29){\footnotesize DD}
 \put(-0.32\linewidth,29){\footnotesize ZNE}
 \put(-0.09\linewidth,29){\footnotesize DD+ZNE}
 \caption{%
 Noise simulation and mitigation of DC-QAOA for portfolio optimization with five qubits using AOQMAP, Qiskit, and Tket. Approximation ratio under (a) depolarizing noise and (b) depolarizing and thermal relaxation noise. We compare four situations from left to right: no error mitigation, applying DD, applying ZNE, and applying both DD and ZNE.
 }\label{fig:ibmq_ehningen_noise_sim}
\end{figure*}

\subsection{Noise simulation and error mitigation}

To assess the impact of noise on the performance of mapped circuits presented in Sec.~\ref{subsec:circu_proper}, we simulate their behavior under noise models. We also explore the effectiveness of various error mitigation strategies in suppressing these noise effects. Appendix~\ref{app:error_mitigation} provides detailed information on strategies including readout error mitigation (REM), dynamical decoupling (DD), and zero-noise extrapolation (ZNE).

We initially employ a widely used depolarizing noise model that introduces random single qubit bit flip, phase flip, and combined bit and phase flip errors into each gate, effectively capturing the effects of mixed noise processes in quantum systems. The depolarizing error channel, denoted as $E_D$, acting on qubits described by density matrix ${\rho}$, is defined as
\begin{equation}
    {E_D}({\rho}) = \frac{\varepsilon }{4}\left({X\rho X} + {Y\rho Y} + {Z\rho Z}\right)+(1 - \frac{3\varepsilon }{4}){\rho},
\end{equation}
where ${X}$, ${Y}$, and ${Z}$ are Pauli matrices, and $\varepsilon $ represents noise parameter.
In addition to depolarizing noise, we also consider thermal relaxation noise \cite{georgopoulos2021modeling}, which arises from the interaction between physical qubits and environment. The device's calibration data is incorporated to evaluate algorithm performance under realistic noise levels. This dataset encompasses essential qubit and gate properties, such as error rates, energy relaxation time $T_1$, and dephasing time $T_2$.

Figure~\ref{fig:ibmq_ehningen_noise_sim} presents simulation results of DC-QAOA for portfolio optimization with five qubits and depths ranging from 1 to 3 on ibmq\_ehningen. We begin by simulating the performance under depolarizing noise and subsequently applying error mitigation techniques including DD and ZNE. In Fig.~\ref{fig:ibmq_ehningen_noise_sim}(a), we present the results without error mitigation, with DD, with ZNE, and with a combination of DD and ZNE. We observe that ZNE proves to be more effective than DD in mitigating errors from depolarizing noise. When both DD and ZNE are applied, the insertion of DD sequences introduces noise, resulting in a slight reduction in approximation ratio. Furthermore, Fig.~\ref{fig:ibmq_ehningen_noise_sim}(b) depicts results for the combination of depolarizing and thermal relaxation noise. While the application of DD in Qiskit enhances approximation ratio of DC-QAOA at high depth ($p=3$), it conversely diminishes the approximation ratio at low depth ($p=1$). This observation highlights a trade-off between mitigating decoherence errors caused by thermal relaxation noise and the introduction gate errors due to insertion of pulse sequences from DD strategy. Additionally, the results demonstrate that a combination of DD and ZNE strategies yields the highest performance.

These findings provide valuable insights into the effectiveness of error mitigation techniques in enhancing the performance of DC-QAOA on QPUs. ZNE proves to be more efficient than DD in mitigating gate errors resulting from depolarizing noise. Moreover, the study highlights the importance of minimizing CX gates and circuit depth to achieve improved performance, as evidenced by higher approximation ratios of AOQMAP, emphasizing the critical role of optimizing circuit properties during compilation process.

\subsection{Demonstration on IBM QPUs}

Following simulation results presented in the previous section, we now demonstrate the performance of DC-QAOA and error mitigation strategies on IBM QPUs, aiming to validate the effectiveness of our optimization strategies in achieving high performance on near-term quantum devices.

\subsubsection{MaxCut on complete graphs}

\begin{figure*}
\centering
	\includegraphics[width=\linewidth]{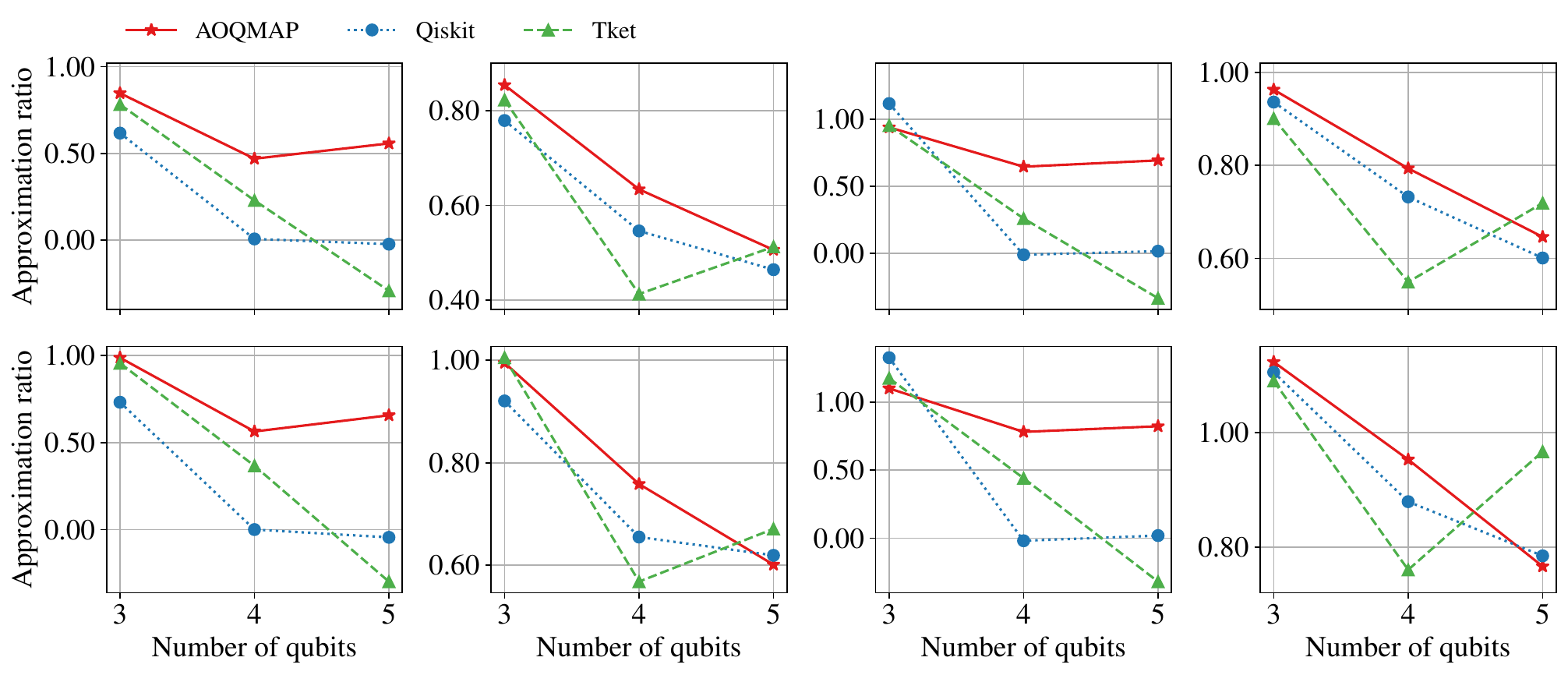}
 \put(-\linewidth,0.4\linewidth){\textbf{(a)}}
 \put(-\linewidth,0.216\linewidth){\textbf{(b)}}
 \put(-0.81\linewidth,0.37\linewidth){\footnotesize Default}
 \put(-0.535\linewidth,0.37\linewidth){\footnotesize DD}
 \put(-0.30\linewidth,0.37\linewidth){\footnotesize ZNE}
 \put(-0.09\linewidth,0.37\linewidth){\footnotesize DD+ZNE}
 \put(-0.79\linewidth,0.19\linewidth){\footnotesize REM}
 \put(-0.585\linewidth,0.19\linewidth){\footnotesize REM+DD}
 \put(-0.350\linewidth,0.19\linewidth){\footnotesize REM+ZNE}
 \put(-0.147\linewidth,0.19\linewidth){\footnotesize REM+DD+ZNE}
 \caption{%
 Approximation ratio of DC-QAOA for MaxCut on complete graphs using AOQMAP, Qiskit, and Tket on ibm\_perth. (a) Results from left to right: no error mitigation, applying DD, applying ZNE, and applying both DD and ZNE. (b) Results corresponding to (a) and additionally applying REM.
 }\label{fig:ibm_perth_maxcut}
\end{figure*}
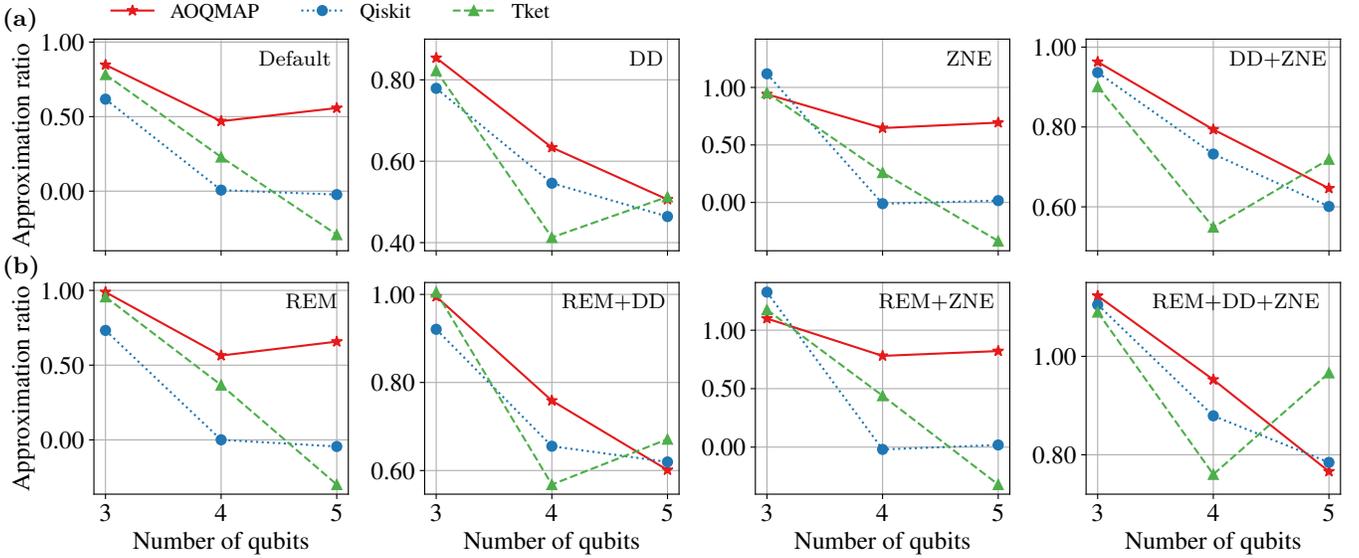

We begin by investigating the MaxCut problem on complete graphs, focusing on evaluating DC-QAOA on QPUs ibm\_perth and ibmq\_ehningen. For noiseless simulations, both DC-QAOA and DC-QAOA-OPT achieve an approximation ratio of 1.00 for tested problem instances, serving as a baseline for their subsequent implementation on QPUs. During execution, we set the number of shots to 50000 for ibm\_perth and 20000 for ibmq\_ehningen, ensuring sufficient coverage. In the ZNE error mitigation process, we use scale factors of \{1, 2, 3\} for ibm\_perth and \{1, 1.5, 2, 2.5, 3\} for ibmq\_ehningen.

\begin{figure*}[ht]
\centering
	\includegraphics[width=\linewidth]{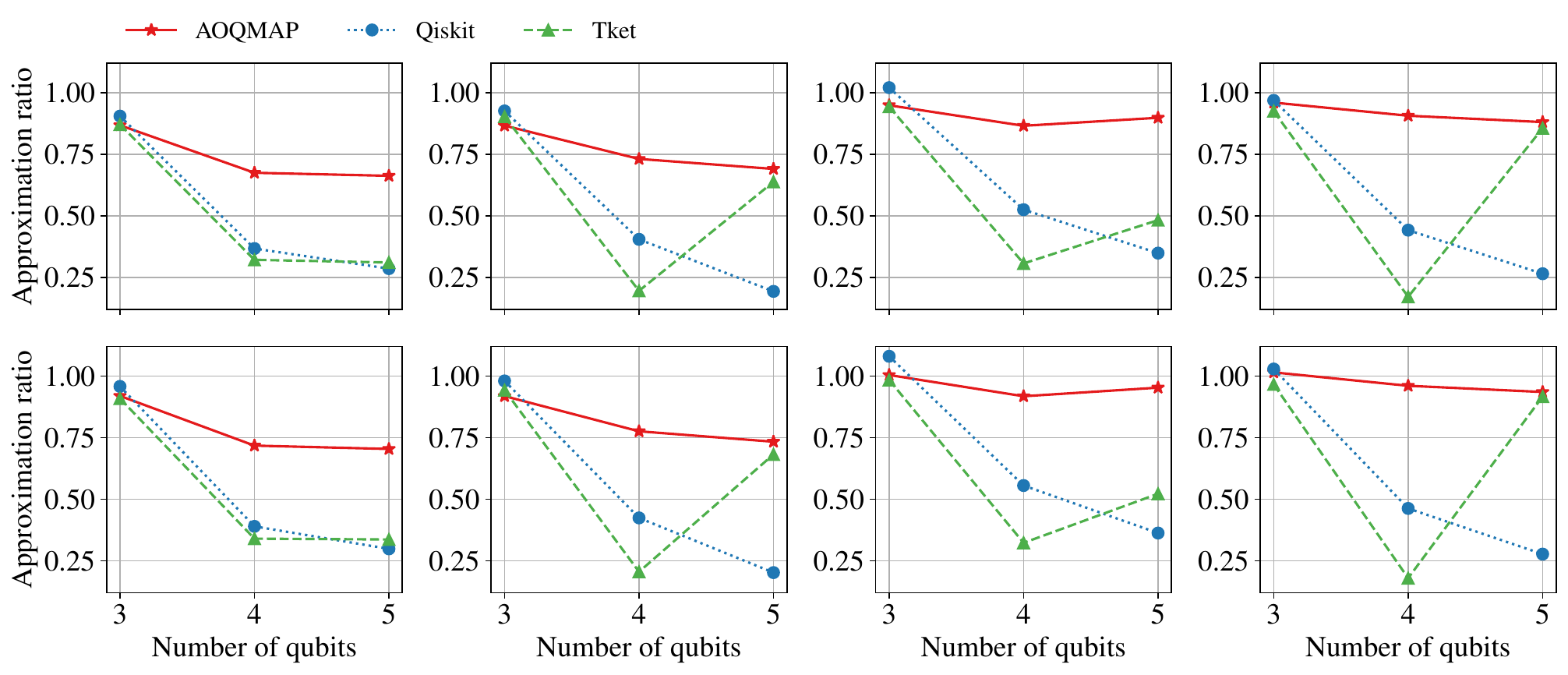}
 \put(-\linewidth,0.4\linewidth){\textbf{(a)}}
 \put(-\linewidth,0.216\linewidth){\textbf{(b)}}
 \put(-0.81\linewidth,0.37\linewidth){\footnotesize Default}
 \put(-0.535\linewidth,0.37\linewidth){\footnotesize DD}
 \put(-0.30\linewidth,0.37\linewidth){\footnotesize ZNE}
 \put(-0.09\linewidth,0.37\linewidth){\footnotesize DD+ZNE}
 \put(-0.79\linewidth,0.19\linewidth){\footnotesize REM}
 \put(-0.585\linewidth,0.19\linewidth){\footnotesize REM+DD}
 \put(-0.350\linewidth,0.19\linewidth){\footnotesize REM+ZNE}
 \put(-0.147\linewidth,0.19\linewidth){\footnotesize REM+DD+ZNE}
 \caption{%
 Demonstration on ibmq\_ehningen. Same as in Fig.~\ref{fig:ibm_perth_maxcut}.
 }\label{fig:ibmq_ehningen_maxcut}
\end{figure*}
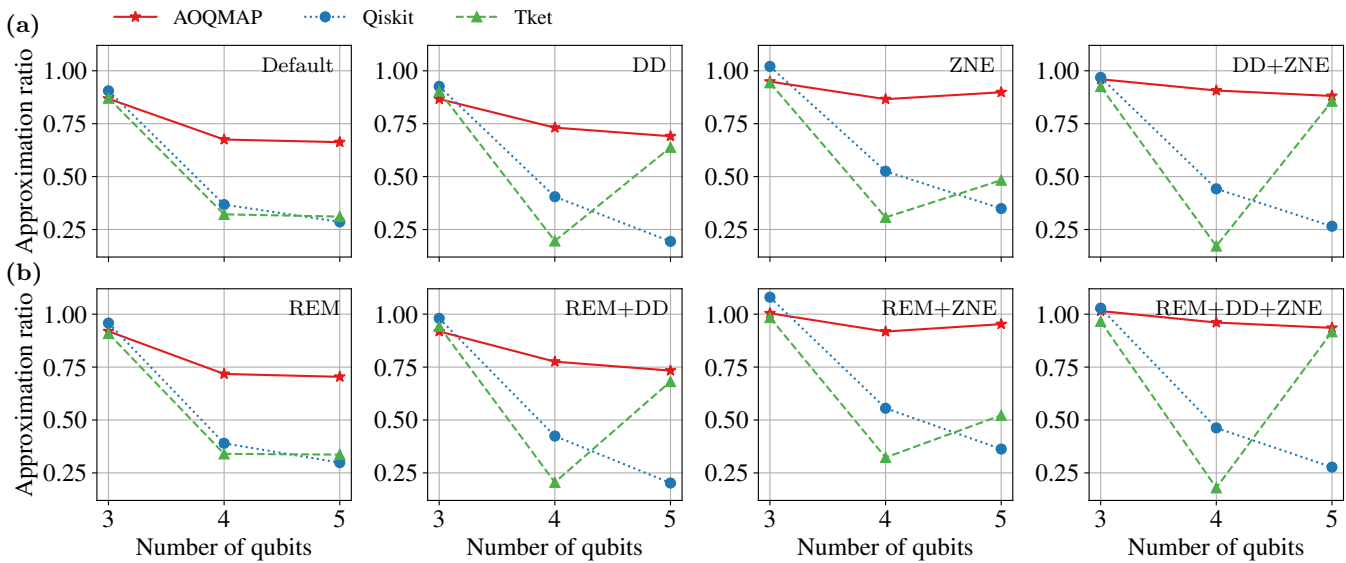

Figure~\ref{fig:ibm_perth_maxcut} presents the results of DC-QAOA and DC-QAOA-OPT on ibm\_perth. Specifically, Figs.~\ref{fig:ibm_perth_maxcut}(a) and \ref{fig:ibm_perth_maxcut}(b) illustrate the approximation ratio obtained without and with REM, respectively. The data show that the application of REM leads to a generally enhanced performance, while the relative behavior of different qubit mapping strategies remains consistent.
AOQMAP achieves overall the highest approximation ratio across all situations, demonstrating its effectiveness with and without applying error mitigation. In comparison, for Qiskit at four and five qubits (Fig.~\ref{fig:ibm_perth_maxcut}(a)), applying DD results in a substantial improvement in approximation ratio, increasing it from zero to around 0.5, while ZNE does not exhibit any noticeable impact. A similar trend is observed for Tket at five qubits, highlighting the effectiveness of DD in mitigating decoherence errors. In contrast, ZNE proves beneficial in improving the performance of AOQMAP with the shortest circuit depth, indicating its value in situations where circuit optimization is prioritized.
The combination of DD and ZNE yields the highest overall performance. Incorporating REM further enhances performance. However, for the three qubits case, we observe that the approximation ratio exceeds one when both ZNE and REM are employed. This discrepancy may arise from potential over-mitigation during the postprocessing stage of ZNE and REM techniques.

Figure~\ref{fig:ibmq_ehningen_maxcut} illustrates results obtained on ibmq\_ehningen using same problem instances as in Fig.~\ref{fig:ibm_perth_maxcut}. AOQMAP consistently outperforms Qiskit and Tket, achieving the highest approximation ratio when combining DD, ZNE, and REM. Similar to observations presented in Fig.~\ref{fig:ibm_perth_maxcut}, applying REM enhances approximation ratios, while the relative performance across various qubit mapping approaches remains unchanged. Furthermore, DD demonstrates significant effectiveness for Tket on five-qubit instances. A potential explanation is that qubits utilized in the execution are particularly sensitive to decoherence noise, rendering DD a beneficial strategy in this scenario.

\subsubsection{Portfolio optimization}

\begin{table*}[t]
\caption{\label{tab:portopt_noiseless}%
Approximation ratio of $n$-qubit DC-QAOA and DC-QAOA-OPT with a depth ranging from 1 to 3 for portfolio optimization in the absence of noise using the QASM simulator.
}
\begin{ruledtabular}
\begin{tabular}{lccc}
\multicolumn{1}{c}{$n$}&
\multicolumn{1}{c}{3}&
\multicolumn{1}{c}{4} &
\multicolumn{1}{c}{5}\\
\colrule
DC-QAOA & [0.985, 0.997, 0.997] & [0.575, 0.821, 0.845] & [0.428, 0.718, 0.829]\\
DC-QAOA-OPT & [0.987, 0.998, 0.998] & [0.739, 0.924, 0.919] & [0.615, 0.882, 0.946]
\end{tabular}
\end{ruledtabular}
\end{table*}

\begin{figure*}
\centering
	\includegraphics[width=\linewidth]{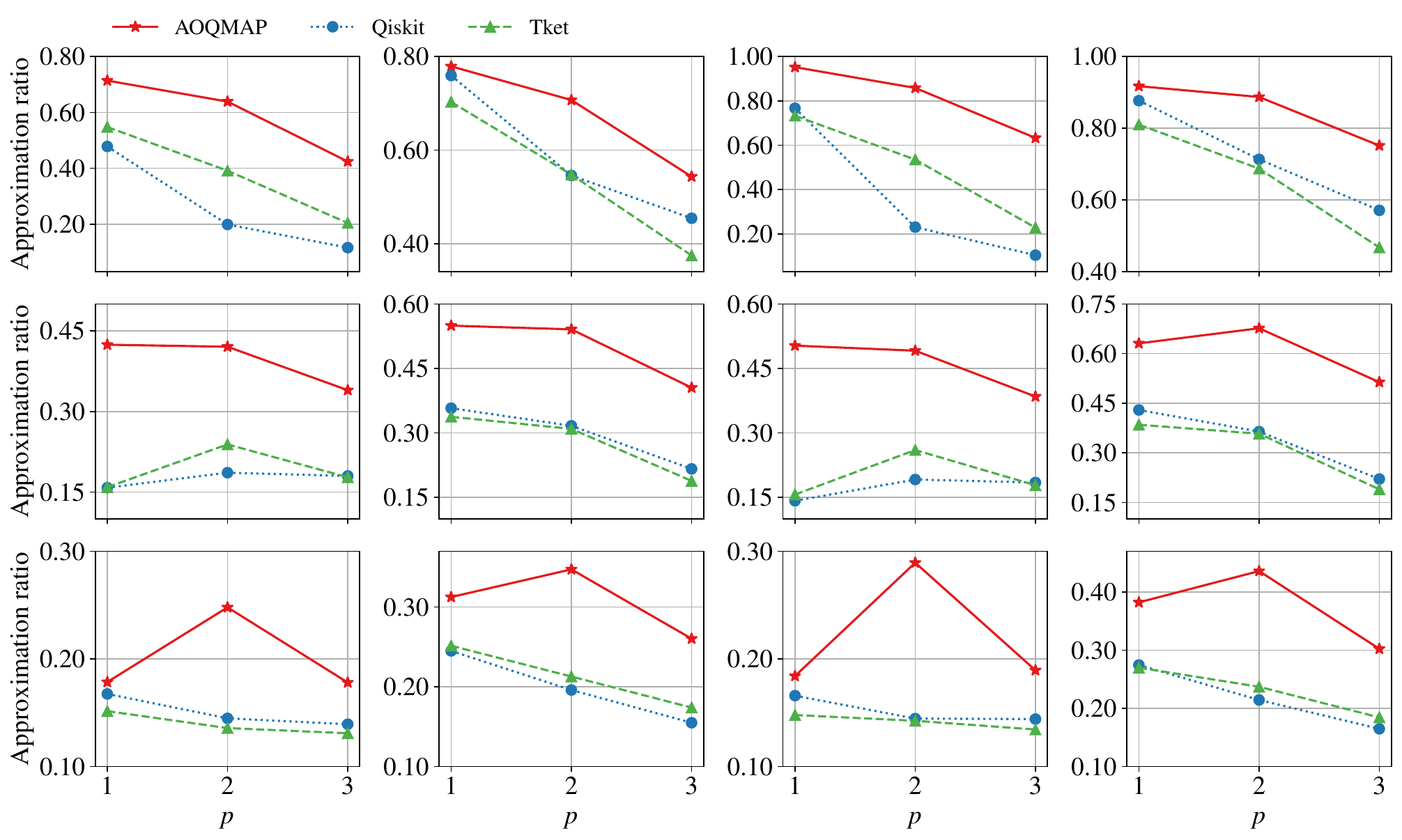}
 \put(-\linewidth,0.57\linewidth){\textbf{(a)}}
 \put(-\linewidth,0.39\linewidth){\textbf{(b)}}
 \put(-\linewidth,0.21\linewidth){\textbf{(c)}}
 \put(-0.81\linewidth,0.54\linewidth){\footnotesize Default}
 \put(-0.535\linewidth,0.54\linewidth){\footnotesize DD}
 \put(-0.30\linewidth,0.54\linewidth){\footnotesize ZNE}
 \put(-0.09\linewidth,0.54\linewidth){\footnotesize DD+ZNE}
 \put(-0.81\linewidth,0.365\linewidth){\footnotesize Default}
 \put(-0.535\linewidth,0.365\linewidth){\footnotesize DD}
 \put(-0.30\linewidth,0.365\linewidth){\footnotesize ZNE}
 \put(-0.09\linewidth,0.365\linewidth){\footnotesize DD+ZNE}
 \put(-0.81\linewidth,0.19\linewidth){\footnotesize Default}
 \put(-0.535\linewidth,0.19\linewidth){\footnotesize DD}
 \put(-0.30\linewidth,0.19\linewidth){\footnotesize ZNE}
 \put(-0.09\linewidth,0.19\linewidth){\footnotesize DD+ZNE}
 \caption{%
Approximation ratio of DC-QAOA for portfolio optimization on ibm\_cairo with (a) three qubits, (b) four qubits, and (c) five qubits. From left to right: no error mitigation, applying DD, applying ZNE, and applying both DD and ZNE.
 }\label{fig:ibm_cairo_portopt}
\end{figure*}

\begin{figure*}
\centering
	\includegraphics[width=\linewidth]{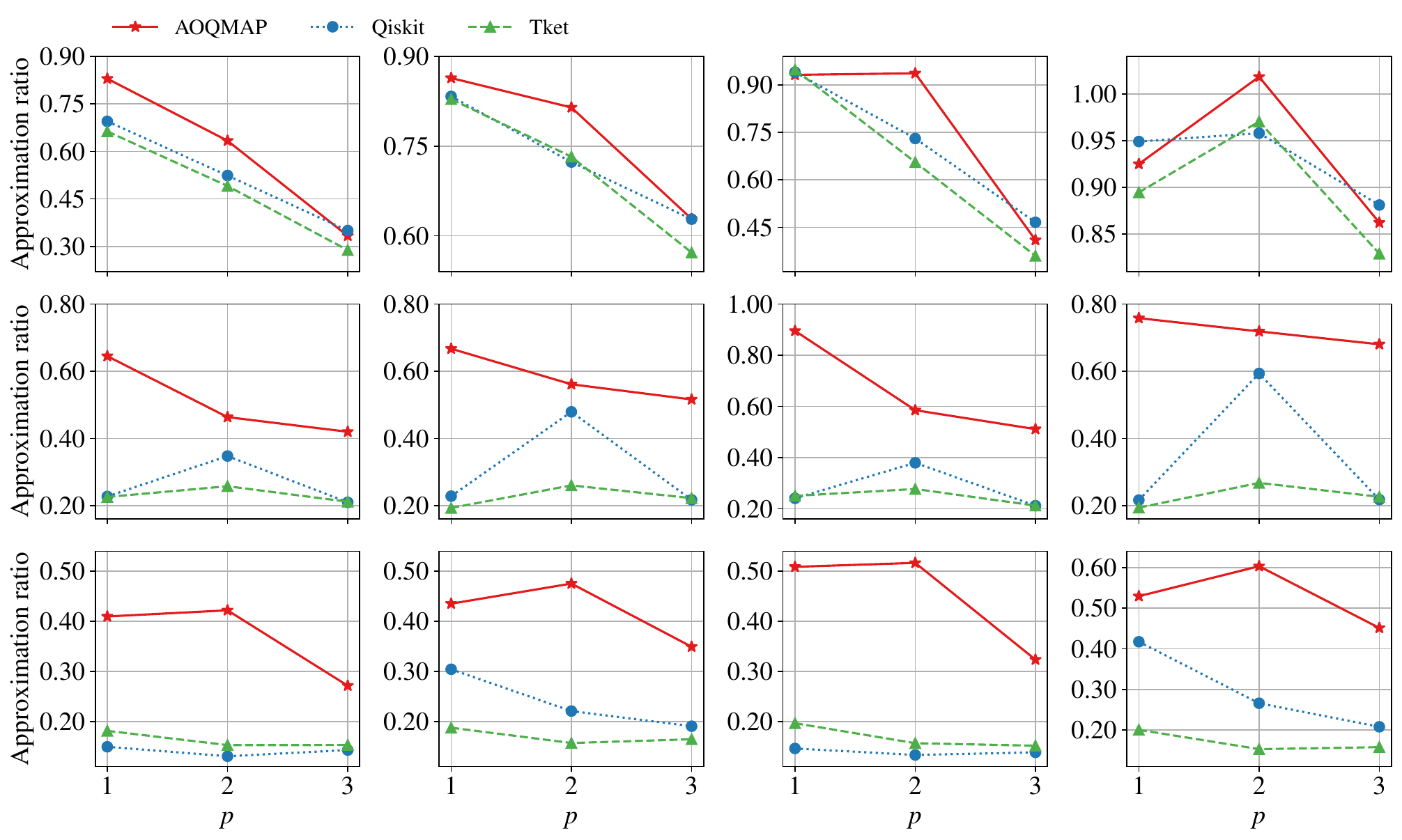}
 \put(-\linewidth,0.57\linewidth){\textbf{(a)}}
 \put(-\linewidth,0.39\linewidth){\textbf{(b)}}
 \put(-\linewidth,0.21\linewidth){\textbf{(c)}}
 \put(-0.81\linewidth,0.54\linewidth){\footnotesize Default}
 \put(-0.535\linewidth,0.54\linewidth){\footnotesize DD}
 \put(-0.30\linewidth,0.54\linewidth){\footnotesize ZNE}
 \put(-0.09\linewidth,0.54\linewidth){\footnotesize DD+ZNE}
 \put(-0.81\linewidth,0.365\linewidth){\footnotesize Default}
 \put(-0.535\linewidth,0.365\linewidth){\footnotesize DD}
 \put(-0.30\linewidth,0.365\linewidth){\footnotesize ZNE}
 \put(-0.09\linewidth,0.365\linewidth){\footnotesize DD+ZNE}
 \put(-0.81\linewidth,0.19\linewidth){\footnotesize Default}
 \put(-0.535\linewidth,0.19\linewidth){\footnotesize DD}
 \put(-0.30\linewidth,0.19\linewidth){\footnotesize ZNE}
 \put(-0.09\linewidth,0.19\linewidth){\footnotesize DD+ZNE}
 \caption{%
Demonstration on ibmq\_ehningen. Same as in Fig.~\ref{fig:ibm_cairo_portopt}.
 }\label{fig:ibmq_ehningen_portopt}
\end{figure*}

We now evaluate the performance of DC-QAOA for portfolio optimization on two 27-qubit QPUs: ibm\_cairo and ibmq\_ehningen. The specific instances and parameters for portfolio optimization problems can be found in \cite{ji2023algorithmoriented}.
Table~\ref{tab:portopt_noiseless} presents noiseless simulation results for the default DC-QAOA with ORIG gate sequence and the optimized DC-QAOA-OPT with AOQMAP. DC-QAOA-OPT consistently exhibits an average higher approximation ratio compared to the default DC-QAOA, demonstrating improved performance at algorithmic level.
For the implementation on both QPUs, we configure the number of shots to 20000, which ensures an adequate number of repetitions for statistical sampling on the given problem instances. Additionally, we employ a set of scale factors \{1, 1.5, 2, 2.5\} in ZNE on both QPUs.

Figures~\ref{fig:ibm_cairo_portopt} and \ref{fig:ibmq_ehningen_portopt} display approximation ratio on ibm\_cairo and ibmq\_ehningen, respectively. REM is not employed in this demonstration. As depicted in Fig.~\ref{fig:ibm_cairo_portopt}, AOQMAP consistently outperforms Qiskit and Tket across all problem instances. For three-qubit (Fig.~\ref{fig:ibm_cairo_portopt}(a)), ZNE proves to be more efficient than DD for AOQMAP, while DD performs better for larger circuit depths produced by Qiskit and Tket. The combination of DD and ZNE yields better results compared to other cases. For four-qubit (Fig.~\ref{fig:ibm_cairo_portopt}(b)), DD demonstrates better performance than ZNE across all qubit mapping strategies. Combining DD and ZNE leads to a significant enhancement in AOQMAP across all depths, with a particularly notable improvement at shallower depths for Tket and Qiskit. A similar trend is observed for five-qubit case (Fig.~\ref{fig:ibm_cairo_portopt}(c)). The results highlight the effectiveness of DD in mitigating decoherence errors for deeper circuits and ZNE in mitigating gate errors for shallower circuits.

On ibmq\_ehningen, for three-qubit (Fig.~\ref{fig:ibmq_ehningen_portopt}(a)), ZNE outperforms DD at low depths, while DD performs better at high depths. Combining DD and ZNE achieves the highest performance.
For four-qubit (Fig.~\ref{fig:ibmq_ehningen_portopt}(b)), all mitigation strategies improve the performance of AOQMAP at each depth. DD demonstrates effectiveness at $p=2$ for Qiskit, while the impact of error mitigation on Tket is not significant. For five-qubit (Fig.~\ref{fig:ibmq_ehningen_portopt}(c)), all strategies also significantly improve the performance of AOQMAP. DD proves to be effective for Qiskit and Tket exhibits no significant changes under error mitigation.

Our findings demonstrate that while error mitigation techniques can enhance algorithm performance on near-term quantum devices, optimizing circuits for reduced depth and two-qubit gate count is essential to effectively leverage these techniques. Moreover, combining different error mitigation techniques presents an efficient approach to further enhance performance. DD proves to be particularly effective for longer circuit depths, whereas ZNE is more beneficial for shorter circuit depths. The combination of DD and ZNE yields consistent improvements in performance across the board, highlighting the synergistic benefits of employing multiple error mitigation techniques. 

\subsubsection{Result analysis}

\begin{table}
\caption{\label{tab:res_analy_em}%
Average increase in approximation ratio with error mitigation applied compared to the case without error mitigation.
}
\begin{ruledtabular}
\begin{tabular}{lcccc}
 &
\multicolumn{1}{c}{\textrm{REM}}&
\multicolumn{1}{c}{\textrm{DD}}&
\multicolumn{1}{c}{\textrm{ZNE}} &
\multicolumn{1}{c}{\textrm{DD+ZNE}}\\
\colrule
AOQMAP & 12.2\% & 22.6\% & 23.9\% & 51.9\%\\
Qiskit & 6.8\% & 5.65$\times$ & 18.7\% & 7.62$\times$\\
Tket & 17.8\% & 2.42$\times$ & 13.4\% & 3.48$\times$\\
\colrule
Mean & 12.3\% & 2.77$\times$ & 18.7\% & 3.87$\times$\\
\end{tabular}
\end{ruledtabular}

\caption{\label{tab:res_analy}%
Average increase in the approximation ratio achieved with AOQMAP compared to Qiskit and Tket.
}
\begin{ruledtabular}
\begin{tabular}{lcccc}
 &
\multicolumn{1}{c}{\textrm{Default}}&
\multicolumn{1}{c}{\textrm{DD}}&
\multicolumn{1}{c}{\textrm{ZNE}} &
\multicolumn{1}{c}{\textrm{DD+ZNE}}\\
\colrule
Qiskit & 5.95$\times$ & 58.6\% & 5.59$\times$ & 69.3\%\\
Tket & 3.04$\times$ & 73.9\% & 3.83$\times$ & 100.4\%\\
\colrule
Mean & 4.49$\times$ & 66.3\% & 4.71$\times$ & 84.8\%\\
\end{tabular}
\end{ruledtabular}
\end{table}

In this section, we present a detailed analysis of the specific achievements of AOQMAP, highlighting its advantages compared to Qiskit and Tket. To ensure a fair comparison, any approximation ratio values below zero were uniformly adjusted to a minimum value of 0.01. This adjustment allows for a consistent and meaningful comparison of the results across different scenarios.

Table~\ref{tab:res_analy_em} summarizes performance improvements obtained by applying error mitigation to AOQMAP, Qiskit, and Tket. The results show that REM yields an average improvement of 12.3\% and is more effective for Tket and AOQMAP compared to Qiskit. On the other hand, DD significantly enhances performance, resulting in a 22.6\% increase for AOQMAP, a $2.42 \times$ increase for Tket, and an impressive $5.65 \times$ increase for Qiskit. This discrepancy can be attributed to the deeper circuits generated by Qiskit and Tket. In contrast, ZNE exhibits higher efficacy for AOQMAP, resulting in an average increase of 23.9\% in approximation ratio. The combination of DD and ZNE demonstrates the highest efficiency, which yields a $3.87 \times$ increase.

Table~\ref{tab:res_analy}  presents an average improvement in approximation ratio achieved by AOQMAP compared to Qiskit and Tket across tested QPUs. Without error mitigation (default), AOQMAP demonstrates an average increase of $4.49 \times$ in approximation ratio. When employing ZNE alone, AOQMAP exhibits an average improvement of $4.71 \times$. When only DD is applied, AOQMAP still maintains an increase of 66.3\%. Combining DD with ZNE further enhances the average improvement to 84.8\%.

These results underscore the effectiveness of AOQMAP compared to Qiskit and Tket, both with and without error mitigation.
The higher quality of circuits generated by AOQMAP offers significant benefits when implementing error mitigation strategies, making the combination of DD and ZNE, as well as their separate applications, particularly efficient. In contrast, Qiskit and Tket rely on a single predominant strategy, which in this study is DD.

\section{\label{sec:concl}Discussion and conclusions}

We propose a methodology that integrates algorithm-oriented qubit mapping with optimization strategies to enhance the performance of digitized counterdiabatic quantum optimization algorithms on near-term quantum devices. Our approach optimizes both algorithm design and hardware implementation to achieve significant performance improvements.
The standard QAOA algorithm employs all-to-all commuting ZZ gates, resulting in gate sequences that do not impact its performance. However, DC-QAOA introduces non-commuting gates in counterdiabatic driving term, making it crucial to optimize Hamiltonian and gate sequence for improved performance. Our findings indicate that different gate sequences, despite having the same Trotter error order, can exhibit varying efficacy during parameter optimization.
Based on this observation, we propose an efficient approach that optimizes Hamiltonian, gate sequence, and algorithm parameters while adhering to hardware connectivity constraints. This comprehensive approach leads to performance gains at both algorithmic and hardware levels.

We evaluate the effectiveness of our approach by benchmarking the optimized and mapped gate sequence obtained through AOQMAP with the original sequence mapped using Qiskit and Tket. The results show that our optimized gate sequence leads to improved performance at the algorithmic level compared to the original one. When considering the impact of noise, the quality of qubit mapping becomes crucial. High-quality solutions produced by AOQMAP result in fewer CX gates and shallower depth, leading to lower noise accumulation and improved resilience against errors. Experimental demonstrations on IBM QPUs further validate the effectiveness of our approach. The optimized implementation DC-QAOA-OPT achieves the highest performance compared to the default implementation. Moreover, circuits mapped with AOQMAP benefit from each error mitigation technique, demonstrating efficiency of AOQMAP in leveraging different mitigation approaches, unlike circuits mapped by Qiskit and Tket, which rely on specific strategies.

In summary, our study demonstrates significant performance improvements achieved through the combination of algorithm-oriented qubit mapping and algorithmic level optimization for digitized counterdiabatic quantum algorithms on near-term quantum devices. The co-optimization of qubit mapping, gate sequences, and algorithm parameters effectively reduces inherent errors, thus enhancing overall performance. These strategies present promising opportunities for addressing real-world optimization problems using NISQ devices. Future research can explore the extension of this cooptimization approach to other quantum algorithms. Moreover, the development of algorithms that strike a balance between optimizing gate sequences and minimizing CX gates for partially connected two-qubit Hamiltonians holds value.

\begin{acknowledgments}

The authors would like to thank Xi Chen, Yue Ban, Koushik Paul, Pranav Chandarana, Thomas Wellens, Andreas Sturm, and Juli\'an Ferreiro-V\'elez for their useful discussions.
We extend special thanks to Koushik for the referenced code that was partially utilized in this research.
We acknowledge the use of IBM Quantum services for this work and to advanced services provided by the IBM Quantum Researchers Program. The views expressed are those of the authors, and do not reflect the official policy or position of IBM or the IBM Quantum team.
This work was supported in part by the Ministry of Economic Affairs, Labour and Tourism Baden Württemberg in the frame of the Competence Center Quantum Computing Baden-Württemberg (project ``QORA").

\end{acknowledgments}

\appendix

\section{Cloud platform details\label{app:cloud_platf}}

\begin{figure*}
\centering
\includegraphics[width=0.27\linewidth]{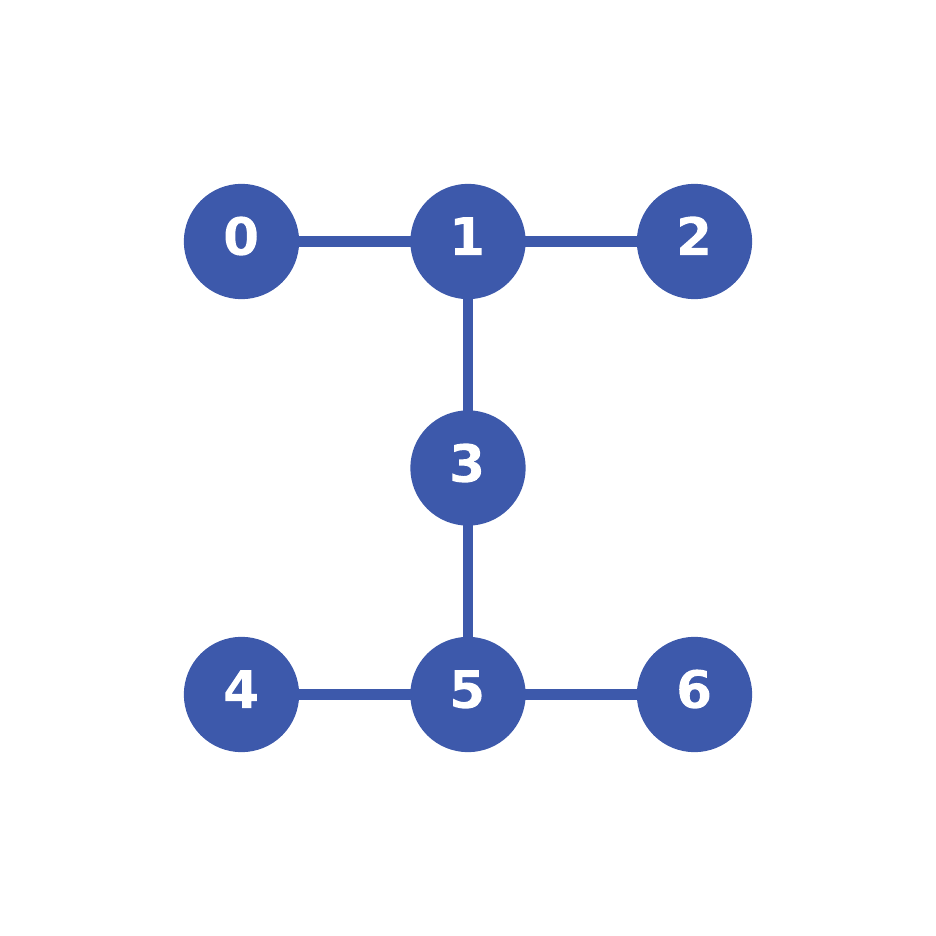}\hspace{5mm}
\includegraphics[width=0.58\linewidth]{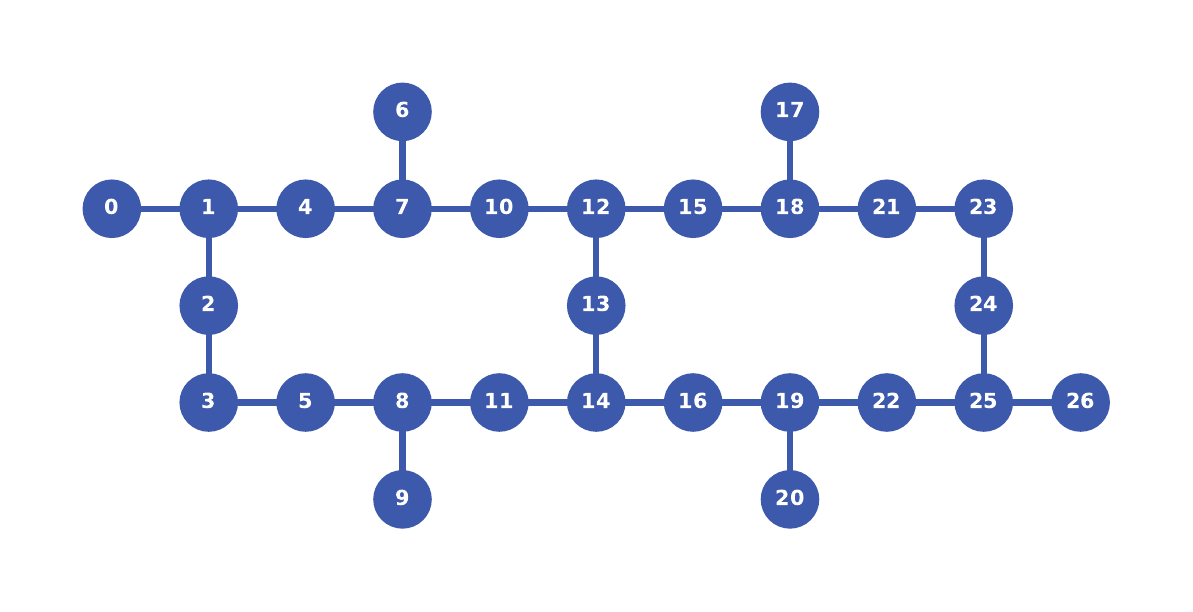}\\
 \caption{%
 IBM QPU topology with 7 (left) and 27 (right) qubits. Each qubit is denoted by a circle, and the lines connecting the qubits depict the available two-qubit gates.
 }\label{fig:gate_maps}
\end{figure*}

Here we provide details on the IBM QPUs employed in this study, including ibm\_perth with 7 qubits, ibm\_cairo, and ibmq\_ehningen, both with 27 qubits. Figure~\ref{fig:gate_maps} shows the topologies of 7-qubit and 27-qubit QPUs. All QPUs employ a basis gate set of \{CX, ID, RZ, SX, X\}. The calibration data at the time of the demonstration are summarized in the following tables. In particular, Tables~\ref{tab:qubits_used_maxcut_complete} and \ref{tab:properties_maxcut_complete} present the qubits employed in the implementation of DC-QAOA for MaxCut problem on ibm\_perth, along with their corresponding qubit properties. Similarly, Tables~\ref{tab:qubits_used_maxcut_complete_ehningen} and \ref{tab:properties_maxcut_complete_ehningen} provide the same information for the MaxCut problem instances executed on ibmq\_ehningen. Furthermore, Tables~\ref{tab:qubits_used_portopt_cairo} and \ref{tab:properties_portopt_cairo} present the qubits utilized and their properties in the implementation of DC-QAOA for portfolio optimization on ibm\_cairo, covering depths ranging from 1 to 3. Finally, Tables~\ref{tab:qubits_used_portopt_ehningen} and \ref{tab:properties_portopt_ehningen} display the corresponding details for the portfolio optimization instances executed on ibmq\_ehningen.

\begin{table*}
\caption{\label{tab:qubits_used_maxcut_complete}%
Qubits used in the $n$-qubit DC-QAOA implementation on ibm\_perth in Fig.~\ref{fig:ibm_perth_maxcut}.
}
\begin{ruledtabular}
\begin{tabular}{lccc}
$n$ &
\multicolumn{1}{c}{\textrm{AOQMAP}}&
\multicolumn{1}{c}{\textrm{Qiskit}}&
\multicolumn{1}{c}{\textrm{Tket}}\\
\colrule
3 & \{1, 2, 3\} & \{1, 3, 5\} & \{1, 2, 3\}\\
4 & \{1, 2, 3, 5\} & \{1, 2, 3, 5\} & \{3, 4, 5, 6] \\
5 & \{1, 2, 3, 5, 6\} & \{1, 3, 4, 5, 6\} & \{1, 3, 4, 5, 6\}\\
\end{tabular}
\end{ruledtabular}
\end{table*}

\begin{table*}
\caption{\label{tab:properties_maxcut_complete}%
Calibration data at the time of the demonstration on ibm\_perth in Fig.~\ref{fig:ibm_perth_maxcut}.
}
\begin{ruledtabular}
\begin{tabular}{lccccccc}
Qubits & 0 & 1 & 2 & 3 & 4 & 5 & 6\\
\colrule
$T_1~(\rm{\mu s})$ & 119.57 & 172.46 & 78.2 & 223.51 & 229.34 & 208.77 & 123.57\\
$T_2~(\rm{\mu s})$ & 68.66 & 51.34 & 98.23 & 292.47 & 145.49 & 236.87 & 238.67\\
Frequency (GHz) & 5.158 & 5.034 & 4.863 & 5.125 & 5.159 & 4.979 & 5.157\\
Anharmonicity (GHz) & $-0.3415$ & $-0.3444$ & $-0.3473$ & $-0.3404$ & $-0.3334$ & $-0.346$ & $-0.3405$\\
Prob meas0 prep1 (\%) & 8.9 & 3.08 & 2.6 & 1.78 & 16.58 & 3.02 & 1.2\\
Prob meas1 prep0 (\%) & 67.38 & 3.16 & 1.88 & 1.62 & 11.94 & 3.02 & 1.04\\
Readout length (ns) & 721.78 & 721.78 & 721.78 & 721.78 & 721.78 & 721.78 & 721.78\\
Readout error (\%) & 38.14 & 3.12 & 2.24 & 1.7 & 14.26 & 3.02 & 1.12\\
Single-qubit gate error (\%) & 0.093 & 0.026 & 0.023 & 0.02 & 0.039 & 0.035 & 0.037\\
CX gate & (0, 1) & (1, 2) & (1, 3) & (3, 5) & (4, 5) & (5, 6)\\
\colrule
CX gate error (\%) & 1.48 & 0.75 & 0.45 & 0.76 & 0.89 & 1.05\\
\end{tabular}
\end{ruledtabular}
\end{table*}

\begin{table*}
\caption{\label{tab:qubits_used_maxcut_complete_ehningen}%
Qubits used in the $n$-qubit DC-QAOA implementation on ibmq\_ehningen in Fig.~\ref{fig:ibmq_ehningen_maxcut}.
}
\begin{ruledtabular}
\begin{tabular}{lccc}
$n$ &
\multicolumn{1}{c}{\textrm{AOQMAP}}&
\multicolumn{1}{c}{\textrm{Qiskit}}&
\multicolumn{1}{c}{\textrm{Tket}}\\
\colrule
3 & \{22, 25, 26\} & \{22, 25, 26\} & \{23, 24, 25\}\\
4 & \{19, 22, 25, 26\} & \{22, 24, 25, 26\} & \{22, 24, 25, 26] \\
5 & \{16, 19, 22, 25, 26\} & \{19, 22, 24, 25, 26\} & \{8, 11, 13, 14, 16\}\\
\end{tabular}
\end{ruledtabular}
\end{table*}

\begin{table*}
\caption{\label{tab:properties_maxcut_complete_ehningen}%
Calibration data at the time of the demonstration on ibmq\_ehningen in Fig.~\ref{fig:ibmq_ehningen_maxcut}.
}
\begin{ruledtabular}
\begin{tabular}{lccccccccccc}
Qubits & 8 & 11 & 13 & 14 & 16 & 19 & 22 & 23 & 24 & 25 & 26 \\
\colrule
$T_1~(\rm{\mu s})$ & 147.13 & 139.39 & 243.31 & 118.9 & 183.36 & 128.91 & 156.26 & 247.34 & 144.53 & 300.87 & 123.47\\
$T_2~(\rm{\mu s})$ & 207.99 & 225.82 & 231.39 & 239.27 & 250.25 & 89.89 & 33.4 & 399.76 & 265.3 & 398.15 & 23.63\\
Frequency (GHz) & 5.174 & 5.119 & 4.926 & 5.177 & 5.022 & 4.784 & 4.725 & 4.805 & 5.074 & 4.95 & 5.151\\
Anharmonicity (GHz) & $-0.3399$ & $-0.3405$ & $-0.344$ & $-0.3408$ & $-0.3435$ & $-0.3485$ & $-0.3464$ & $-0.3471$ & $-0.3416$ & $-0.3457$ & $-0.3391$\\
Prob meas0 prep1 (\%) & 1.4 & 3.88 & 1.04 & 1.4 & 1.0 & 1.14 & 1.62 & 1.04 & 1.1 & 0.84 & 1.0\\
Prob meas1 prep0 (\%) & 1.1 & 1.76 & 0.68 & 0.48 & 0.58 & 0.96 & 1.18 & 0.76 & 0.52 & 0.48 & 0.6\\
Readout length (ns) & 846.22 & 846.22 & 846.22 & 846.22 & 846.22 & 846.22 & 846.22 & 846.22 & 846.22 & 846.22 & 846.22\\
Readout error (\%) & 1.25 & 2.82 & 0.86 & 0.94 & 0.79 & 1.05 & 1.4 & 0.9 & 0.81 & 0.66 & 0.8\\
Single-qubit gate error (\%) & 0.029 & 0.031 & 0.022 & 0.032 & 0.023 & 0.026 & 0.019 & 0.017 & 0.021 & 0.013 & 0.028\\
CX gate & (8, 11) & (11, 14) & (13, 14) & (14, 16) & (16, 19) & (19, 22) & (22, 25) & (23, 24) & (24, 25) & (25, 26)\\
\colrule
CX gate error (\%) & 1.4 & 0.81 & 0.72 & 0.97 & 0.52 & 0.48 & 0.36 & 0.38 & 0.88 & 0.52\\
\end{tabular}
\end{ruledtabular}
\end{table*}

\begin{table*}
\caption{\label{tab:qubits_used_portopt_cairo}%
Qubits used in the $n$-qubit DC-QAOA implementation with depth $p$, represented as $D_n^p$, on ibm\_cairo in Fig.~\ref{fig:ibm_cairo_portopt}.
}
\begin{ruledtabular}
\begin{tabular}{lccc}
 &
\multicolumn{1}{c}{\textrm{AOQMAP}}&
\multicolumn{1}{c}{\textrm{Qiskit}}&
\multicolumn{1}{c}{\textrm{Tket}}\\
\colrule
$D_3^1$ & \{12, 13, 14\} & \{12, 13, 14\} & \{12, 13, 15\}\\
$D_3^2$ & \{3, 5, 8\} & \{12, 13, 14\} & \{12, 13, 15\}\\
$D_3^3$ & \{13, 14, 16\} & \{12, 13, 14\} & \{12, 13, 15\}\\
$D_4^1$ &\{10, 12, 13, 14\} & \{10, 12, 13, 15\} & \{10, 12, 13, 15\}\\
$D_4^2$ & \{3, 5, 8, 9\} & \{10, 12, 13, 15\} & \{10, 12, 13, 15\}\\
$D_4^3$ & \{3, 5, 8, 9\} & \{10, 12, 13, 15\} & \{10, 12, 13, 15\}\\
$D_5^1$ & \{10, 12, 13, 14, 16\} & \{10, 12, 13, 14, 15\} & \{10, 12, 13, 14, 15\}\\
$D_5^2$ & \{10, 12, 13, 14, 16\} & \{10, 12, 13, 14, 15\} & \{10, 12, 13, 14, 15\}\\
$D_5^3$ & \{10, 12, 13, 14, 16\} & \{10, 12, 13, 14, 15\} & \{10, 12, 13, 14, 15\}\\
\end{tabular}
\end{ruledtabular}
\end{table*}

\begin{table*}
\caption{\label{tab:properties_portopt_cairo}%
Calibration data at the time of the demonstration on ibm\_cairo in Fig.~\ref{fig:ibm_cairo_portopt}.
}
\begin{ruledtabular}
\begin{tabular}{lcccccccccc}
Qubits & 3 & 5 & 8 & 9 & 10 & 12 & 13 & 14 & 15 & 16\\
\colrule
$T_1~(\rm{\mu s})$ & 89.65 & 105.74 & 83.87 & 77.23 & 95.75 & 99.99 & 75.72 & 77.0 & 130.63 & 98.95\\
$T_2~(\rm{\mu s})$ & 134.52 & 50.84 & 43.33 & 30.47 & 13.92 & 102.35 & 155.06 & 132.19 & 334.49 & 207.13\\
Frequency (GHz) & 5.119 & 5.046 & 4.969 & 5.227 & 5.234 & 5.115 & 5.282 & 5.044 & 4.963 & 5.277\\
Anharmonicity (GHz) & $-0.3404$ & $-0.3047$ & $-0.3442$ & $-0.3391$ & $-0.3392$ & $-0.3413$ & $-0.3686$ & $-0.3422$ & $-0.3453$ & $-0.3388$\\
Prob meas0 prep1 (\%) & 1.72 & 1.82 & 2.04 & 2.76 & 1.0 & 1.02 & 0.5 & 1.04 & 0.78 & 1.92\\
Prob meas1 prep0 (\%) & 1.72 & 1.36 & 2.6 & 5.48 & 0.8 & 0.98 & 0.36 & 0.58 & 0.42 & 10.18\\
Readout length (ns) & 732.44 & 732.44 & 732.44 & 732.44 & 732.44 & 732.44 & 732.44 & 732.44 & 732.44 & 732.44\\
Readout error (\%) & 1.72 & 1.59 & 2.32 & 4.12 & 0.9 & 1.0 & 0.43 & 0.81 & 0.6 & 6.05\\
Single-qubit gate error (\%) & 0.02 & 0.024 & 0.016 & 0.022 & 0.019 & 0.019 & 0.015 & 0.02 & 0.011 & 0.02\\
CX gate & (3, 5) & (5, 8) & (8, 9) & (10, 12) & (12, 13) & (12, 15) & (13, 14) & (14, 16)\\
\colrule
CX gate error (\%) & 0.55 & 0.54 & 0.8 & 0.8 & 1.16 & 0.93 & 0.48 & 0.52\\
\end{tabular}
\end{ruledtabular}

\caption{\label{tab:qubits_used_portopt_ehningen}%
Qubits used in the $n$-qubit DC-QAOA implementation with depth $p$, represented as $D_n^p$, on ibmq\_ehningen in Fig.~\ref{fig:ibmq_ehningen_portopt}.
}
\begin{ruledtabular}
\begin{tabular}{lccc}
 &
\multicolumn{1}{c}{\textrm{AOQMAP}}&
\multicolumn{1}{c}{\textrm{Qiskit}}&
\multicolumn{1}{c}{\textrm{Tket}}\\
\colrule
$D_3^1$ & \{22, 25, 26\} & \{22, 25, 26\} & \{23, 24, 25\}\\
$D_3^2$ & \{19, 22, 25\} & \{22, 25, 26\} & \{23, 24, 25\}\\
$D_3^3$ & \{19, 22, 25\} & \{22, 25, 26\} & \{23, 24, 25\}\\
$D_4^1$ & \{19, 22, 25, 26\} & \{22, 24, 25, 26\} & \{22, 24, 25, 26\}\\
$D_4^2$ & \{19, 22, 25, 26\} & \{19, 22, 25, 26\} & \{22, 24, 25, 26\}\\
$D_4^3$ & \{19, 22, 25, 26\} & \{22, 24, 25, 26\} & \{22, 24, 25, 26\}\\
$D_5^1$ & \{16, 19, 22, 25, 26\} & \{16, 19, 20, 22, 25\} & \{19, 22, 24, 25, 26\}\\
$D_5^2$ & \{16, 19, 22, 25, 26\} & \{16, 19, 20, 22, 25\} & \{19, 22, 24, 25, 26\}\\
$D_5^3$ & \{16, 19, 22, 25, 26\} & \{16, 19, 20, 22, 25\} & \{19, 22, 24, 25, 26\}\\
\end{tabular}
\end{ruledtabular}

\caption{\label{tab:properties_portopt_ehningen}%
Calibration data at the time of the demonstration on ibmq\_ehningen in Fig.~\ref{fig:ibmq_ehningen_portopt}.
}
\begin{ruledtabular}
\begin{tabular}{lcccccccc}
Qubits & 16 & 19 & 20 & 22 & 23 & 24 & 25 & 26\\
\colrule
$T_1~(\rm{\mu s})$ & 183.36 & 128.91 & 43.22 & 156.26 & 247.34 & 144.53 & 300.87 & 123.47\\
$T_2~(\rm{\mu s})$ & 250.25 & 89.89 & 121.72 & 33.4 & 399.76 & 265.3 & 398.15 & 23.63\\
Frequency (GHz) & 5.022 & 4.784 & 5.042 & 4.725 & 4.805 & 5.074 & 4.95 & 5.151\\
Anharmonicity (GHz) & $-0.3435$ & $-0.3485$ & $-0.3426$ & $-0.3464$ & $-0.3471$ & $-0.3416$ & $-0.3457$ & $-0.3391$\\
Prob meas0 prep1 (\%) & 1.0 & 1.14 & 0.82 & 1.62 & 1.04 & 1.1 & 0.84 & 1.0\\
Prob meas1 prep0 (\%) & 0.58 & 0.96 & 2.18 & 1.18 & 0.76 & 0.52 & 0.48 & 0.6\\
Readout length (ns) & 846.22 & 846.22 & 846.22 & 846.22 & 846.22 & 846.22 & 846.22 & 846.22\\
Readout error (\%) & 0.79 & 1.05 & 1.5 & 1.4 & 0.9 & 0.81 & 0.66 & 0.8\\
Single-qubit gate error (\%) & 0.023 & 0.026 & 0.038 & 0.019 & 0.017 & 0.021 & 0.013 & 0.028\\
CX gate & (16, 19) & (19, 20) & (19, 22) & (22, 25) & (23, 24) & (24, 25) & (25, 26)\\
\colrule
CX gate error (\%) & 0.52 & 0.57 & 0.48 & 0.36 & 0.38 & 0.88 & 0.52\\
\end{tabular}
\end{ruledtabular}
\end{table*}

\section{Error mitigation techniques\label{app:error_mitigation}}

We introduce several error mitigation techniques \cite{huang2023near}, including readout error mitigation (REM), dynamical decoupling (DD), and zero-noise extrapolation (ZNE). These methods play a crucial role in enhancing the reliability and accuracy of quantum computations, mitigating the effects of errors in different ways.

Readout errors pose a significant challenge in current quantum systems, with error rates typically higher than single qubit gate errors by approximately one order of magnitude. These errors stem from various sources, including decoherence during the measurement process and the overlapping of measured physical quantities associated with $|0\rangle$ and $|1\rangle$ states. A range of mitigation techniques have been proposed \cite{bravyi2021mitigating, geller2021toward, nation2021scalable, maciejewski2021modeling, funcke2022measurement, geller2020rigorous, chen2019detector, van2022model, maciejewski2020mitigation, nachman2020unfolding, hicks2021readout, peters2023perturbative}, aiming to enhance the precision of quantum state measurements and minimize their impact, which typically involve constructing noise models or applying classical postprocessing to measured outcomes. In this study, we investigate the influence of REM on the algorithm performance by employing the matrix-free measurement mitigation (M3) technique \cite{nation2021scalable}. M3 is a scalable method for mitigating measurement errors that eliminates the need to form or invert the entire assignment matrix. Instead, it operates within a subspace defined by the noisy input bit strings.

Dynamical decoupling (DD) \cite{suter2016colloquium, ahmed2013robustness, pokharel2018demonstration, souza2012robust} is a widely used error mitigation technique in quantum systems. Its primary objective is to suppress decoherence errors during qubit idle times by applying a sequence of pulses. DD is effective in protecting quantum states against noise and can also be utilized to mitigate crosstalk \cite{tripathi2022suppression} and coherent errors \cite{qiu2021suppressing}.
DD sequences typically consist of repetitive patterns of pulses, such as $\pi$ pulses, which counteract the detrimental effects of system and environment coupling. The relative phases and delays between pulses play a crucial role in improving the robustness of DD sequences against pulse imperfections and unwanted environmental interactions.
A variety of DD sequences, such as Carr-Purcell-Meiboom-Gill (CPMG) \cite{carr1954effects, meiboom2004modified}, XY4 \cite{maudsley1986modified, alvarez2012iterative, viola1999dynamical, souza2012effects}, KDD \cite{souza2011robust}, and Uhrig dynamical decoupling (UDD) \cite{uhrig2007keeping}, have been investigated and demonstrated promising results in various quantum systems, such as spin systems \cite{merkel2021dynamical}, superconducting qubits \cite{bylander2011noise}, and quantum memories \cite{biercuk2009optimized}. Additionally, DD techniques have been applied in simulations of quantum many-body systems \cite{chen2022error} and quantum machine learning \cite{melo2023pulse}.
In this study, we leverage the CPMG pulse sequence to mitigate errors within the quantum circuit. The CPMG pulse sequence, denoted as $\tau/4$ -- X -- $\tau/2$ -- X -- $\tau/4$, involves applying two X pulses, separated by a delay $\tau/2$, with additional delays of $\tau/4$ at the beginning and end of the pulse sequence. The delay time $\tau$ is the idle time of the qubit minus the duration of two X pulses.

Zero-noise extrapolation (ZNE) estimates ideal expectation values from noisy data by leveraging measurements taken at different controlled noise levels \cite{li2017efficient, temme2017error, endo2018practical, kandala2019error}. It comprises two primary steps: noise scaling and extrapolation. Noise scaling is the process of modifying the quantum circuit to run at different noise levels. This can be achieved through pulse stretching \cite{kandala2019error} or by applying digital techniques such as gate folding \cite{giurgica2020digital}. Extrapolation is the process of fitting a mathematical model to noisy data and estimating the zero-noise value using different methods such as linear, polynomial, exponential, or adaptive extrapolation. 
A similar approach to ZNE is the probabilistic error cancellation (PEC) \cite{temme2017error, endo2018practical, zhang2020error, mari2021extending}, which also aims to reduce the impact of noise by sampling from a collection of circuits that emulate a noise inverting channel, effectively canceling out the noise. In this study, we employ ZNE to mitigate errors, focusing on folding two-qubit gates, particularly CX gates. We utilize the linear fitting method due to its simplicity and effectiveness.

\bibliography{refs}

\end{document}